\documentclass[12pt,preprint]{aastex}

\def\ni{\noindent}
\def\s{{\rm s}}

\def\cm{{\rm\,cm}}
\def\m{{\rm\,m}}
\def\km{{\rm\,km}}

\def\gm{{\rm\,g}}
\def\g{{\rm\,g}}

\def\yr{{\rm\,yr}}

\begin{document}

\shortauthors{Chiang and Lithwick}
\shorttitle{Origin of Neptune Trojans}

\title{Neptune Trojans as a Testbed for Planet Formation}

\author{E. I. Chiang\altaffilmark{1,2} \& Y. Lithwick\altaffilmark{1}}

\altaffiltext{1}{Astronomy Department,
University of California at Berkeley,
Berkeley, CA~94720, USA}

\altaffiltext{2}{Alfred P.~Sloan Research Fellow}

\email{echiang@astron.berkeley.edu}

\begin{abstract}
The problem of accretion in the Trojan 1:1 resonance is akin to the standard
problem of planet formation, transplanted from a star-centered disk
to a disk centered on the Lagrange point.
The newly discovered class of Neptune Trojans
promises to test theories of planet formation by coagulation.
Neptune Trojans resembling the prototype
2001 QR$_{322}$ (``QR'')---whose radius of $\sim$100 km
is comparable to that of the largest Jupiter Trojan---may
outnumber their Jovian counterparts by a factor
of $\sim$10. We develop and test three theories for the origin
of large Neptune Trojans: pull-down
capture, direct collisional emplacement, and {\it in situ} accretion.
These theories are staged
after Neptune's orbit anneals: after dynamical friction
eliminates any large orbital eccentricity and after the planet
ceases to migrate. We discover that seeding the 1:1 resonance
with debris from planetesimal collisions and having the seed
particles accrete {\it in situ} naturally reproduces the inferred
number of QR-sized Trojans.
We analyze accretion in the Trojan sub-disk by applying
the two-groups method, accounting for kinematics specific to
the resonance. We find that a
Trojan sub-disk comprising decimeter-sized seed particles
and having a surface density $\sim$$10^{-3}$ that
of the local minimum-mass disk produces
$\sim$10 QR-sized objects in $\sim$1 Gyr, in accord with observation.
Further growth is halted by collisional diffusion of seed particles
out of resonance. In our picture, the number and sizes of the largest
Neptune Trojans represent the unadulterated outcome
of dispersion-dominated, oligarchic accretion.
Large Neptune Trojans, perhaps the most newly accreted
objects in our Solar System, may today have a dispersion
in orbital inclination of less than $\sim$10 degrees,
despite the existence of niches of stability at higher
inclinations. Such a vertically thin disk, born of a dynamically cold
environment necessary for accretion, and raised in minimal contact
with external perturbations, contrasts with the thick
disks of other minor body belts.
\end{abstract}


\section{INTRODUCTION}
\label{intro}

Trojans are planetesimals that trace tadpole-shaped trajectories around
one of two triangular equilibrium points (Lagrange points)
established by their host planet (Lagrange 1873; Murray \& Dermott 1999).
They are said to inhabit the 1:1 resonance because they execute one
orbit about the Sun for every orbit that their host planet makes, staying
an average of $\sim$60$^{\circ}$ forwards or backwards of the planet's
orbital longitude. The Lagrange points
represent potential maxima in the frame rotating with
the planet's mean orbital frequency.
The Coriolis force renders Trojan motion dynamically stable,
while dissipative forces (such as introduced by inter-Trojan collisions)
that reduce the energy in the rotating frame cause Trojans
to drift from their potential maxima and to
escape the resonance.\footnote{For this reason, Trojan motion
is sometimes referred to as dynamically stable but secularly unstable.
We will account explicitly for various forms of secular instability
experienced by Trojans in \S\ref{insitu}.}
Best known are the Jupiter Trojans:
two clouds of rocky, icy bodies that flank the gas giant
and whose sizes range up to that of (624) Hektor,
which has a characteristic radius $R \sim 100\km$ (Barucci et al.~2002; Marzari
et al.~2002; Jewitt, Sheppard, \& Porco 2004).
The number of kilometer-sized Jupiter Trojans may
exceed that of similarly sized Main Belt asteroids (Jewitt, Trujillo, \& Luu
2000).

Recently, the first Neptune Trojan, 2001 QR$_{322}$ (hereafter
``QR''),  was discovered by the Deep Ecliptic Survey (DES),
an observational reconnaissance of the
Kuiper belt at optical wavelengths (Chiang et al.~2003, hereafter C03).
This Hektor-sized object librates (oscillates) about Neptune's forward
Lagrange (L4) point and vindicates long-standing theoretical beliefs in
the dynamical stability of Neptune Trojans (Holman \& Wisdom 1993;
Nesvorny \& Dones 2002). Billion-year-long orbital
integrations of possible trajectories of QR robustly indicate
stability and suggest that the object has inhabited the 1:1 resonance for the
age of the Solar System, $t_{\rm sol} \approx 4.6 \times 10^9\yr$ (C03;
Marzari, Tricarico, \& Scholl 2003; Brasser et al.~2004).
Extrapolation of the total population of Neptune
Trojans based on the amount of sky surveyed by the DES indicates
that Neptune Trojans resembling QR may be 10--30 times
as numerous as their Jovian counterparts (C03).
If so, their surface mass density would approach
that of the current main Kuiper belt to within a factor of a few
[e.g., Bernstein et al.~2004; see also equations (\ref{gmin1}) and
(\ref{gmin2}) of this paper].

Here we investigate the origin of Neptune Trojans.
Unlike other resonant Kuiper belt objects (3:2, 2:1, 5:2, etc.)
whose existence may be explained by the outward
migration of Neptune in the primordial disk of planetesimals
and concomitant resonance trapping
(Malhotra 1995; Chiang \& Jordan 2002; Murray-Clay \& Chiang 2004),
Neptune Trojans do not owe their genesis to migration.
As a planet migrates
on timescales much longer than the local orbital period, it
scatters neighboring planetesimals onto
extremely elongated and inclined orbits
by repeated close encounters (C03; Gomes 2003).
Such scattering might explain
the large velocity dispersions---eccentricities
and inclinations with respect to the invariable
plane ranging up to $\sim$0.2 and $\sim$0.5 rad,
respectively---observed in the main Kuiper belt today (see, e.g.,
Elliot et al.~2005; Gomes 2003).
By contrast, the orbit of QR is nearly circular and nearly co-planar
with the invariable plane;
its eccentricity and inclination are $\sim$0.03 and $\sim$0.02 rad,
respectively. In simulations of migration and resonance
trapping executed by C03, the sweeping 1:1 resonance fails
to trap a single test particle. Instead, Neptune's migration may
destabilize Neptune Trojans. Passage
of Neptune Trojans through sweeping secondary resonances
with Uranus and the other giant planets
can reduce the Trojan population by nearly 2 orders of magnitude
(Gomes 1998; Kortenkamp, Malhotra, \& Michtchenko 2004).

Do Jupiter Trojans offer any insight into the formation of Neptune
Trojans? Morbidelli et al.~(2004) propose that in the early
planetesimal disk, Jupiter Trojans
are captured as Jupiter and Saturn migrate divergently
across their mutual 2:1 resonance
(see also Chiang 2003 for a more general discussion of divergent
resonance crossings). While Jupiter and Saturn occupy the 2:1 resonance,
Jupiter's 1:1 resonance is unstable;
planetesimals stream through Trojan libration regions
on orbits that tend to be highly inclined due to planetary
perturbations. Once the planets depart the 2:1 resonance,
stability returns to the 1:1 resonance. At this moment,
planetesimals that happen to be passing though
Trojan libration regions are trapped there, effectively instantaneously.
This ``freeze-in'' scenario promises to explain the large orbital
inclinations---up to $\sim$0.6 rad---exhibited by Jupiter
Trojans, in addition to Saturn's observed orbital eccentricity of $\sim$0.05
which might result from the planets' resonance crossing.
While we cannot rule out that Neptune Trojans did not also form
by freeze-in, analogous motivating observations are absent:
The orbital inclination of QR is low;
divergent crossing of the 2:1 resonance by Uranus and Neptune
may cause Neptune's eccentricity to exceed
its observed value of $\sim$0.01 by a factor of $\sim$3; and finally,
the stability of Plutinos and other resonantly trapped
Kuiper belt objects is threatened by planetary resonance
crossing (Gomes 2001).

Another motivation for studying Neptune Trojans is to infer
the formation environment of the host ice giant.
Neptune's formation is the subject of substantial
current research because traditional estimates of the planet's accretion
timescale are untenably longer than $t_{\rm sol}$ (Thommes, Levison, \& Duncan
1999; Goldreich, Lithwick, \& Sari~2004, hereafter GLS04).
As planetesimals that share Neptune's orbit,
Neptune Trojans may hold clues as to how their host planet assembled.
Their composition probably reflects that of Neptune's rock/ice interior.

We quantitatively develop and assess the viability of three theories
for the origin of QR-like objects:

\begin{enumerate}

\item{Pull-down capture, whereby mass accretion of the host planet
converts a planetesimal's orbit into tadpole-type libration.}

\item{Direct collisional emplacement, whereby initially
non-resonant, QR-sized objects are diverted into 1:1
resonance by physical collisions.}

\item{{\it In situ} accretion, whereby QR-sized bodies
form by accretion of much smaller seed particles comprising
a Trojan sub-disk in the solar nebula. Seed particles
are presumed to be inserted into resonance as debris
from collisions between planetesimals. The problem of
accretion in the Trojan sub-disk is akin to the standard
problem of planet formation, transplanted from the usual
heliocentric setting to an L4/L5-centric environment.}

\end{enumerate}

\ni Peale (1993) analyzes trapping of Jupiter Trojans
by nebular gas drag. We do not consider gas dynamics
since the outer Solar System after the time of
Neptune's formation was gas-depleted,
almost certainly due to photo-evaporation
by ultraviolet radiation from the young Sun (e.g., Matsuyama,
Johnstone, \& Hartmann 2003, and references therein).
By mass, Neptune comprises only $\sim$4--18\% hydrogen and helium
(Lodders \& Fegley 1998).

While the three mechanisms we examine are not mutually exclusive,
the requirements and predictions that each makes
independent of the others differ.
Faced with only a single known example of a Neptune Trojan and limited
data concerning its physical properties,
we wield order-of-magnitude physics as our weapon of choice.
Many of our simple estimates prove surprisingly illuminating.

In \S\ref{basic}, we review the collisionless dynamics
of Trojans and supply relations and terminology that will be used
in remaining sections.
Observed and theoretically inferred properties
of Neptune Trojans requiring explanation
are listed in \S\ref{character}. There, we also place the birth
of Neptune Trojans on the time-line of Neptune's formation
and orbital evolution.
In \S\ref{pulldown}, we argue against pull-down capture
as the primary channel for formation.
In \S\ref{deposit}, we quantify and assess
the plausibility of demands that direct collisional emplacement
places on the planetesimal disk. In \S\ref{insitu}, we demonstrate how
{\it in situ} accretion can
correctly reproduce the inferred number of QR-sized Neptune Trojans.
In \S\ref{conclude}, we summarize our findings
and point out directions for future observational and theoretical work.

\section{BASIC TROJAN MOTION}
\label{basic}
We review the motions of Trojans hosted by a planet
on a circular orbit to establish
simple relations used throughout this paper.
Some of the material in this section is derived in standard
textbooks (e.g., Murray \& Dermott 1999).
Exceptions include Trojan-Trojan relative motion
and the variation of libration period
with tadpole size, topics that
we develop ourselves.

\subsection{Epicyclic Motion}
\label{epimotion}
\begin{figure}
\epsscale{1.1}
\vspace{-2in}
\plotone{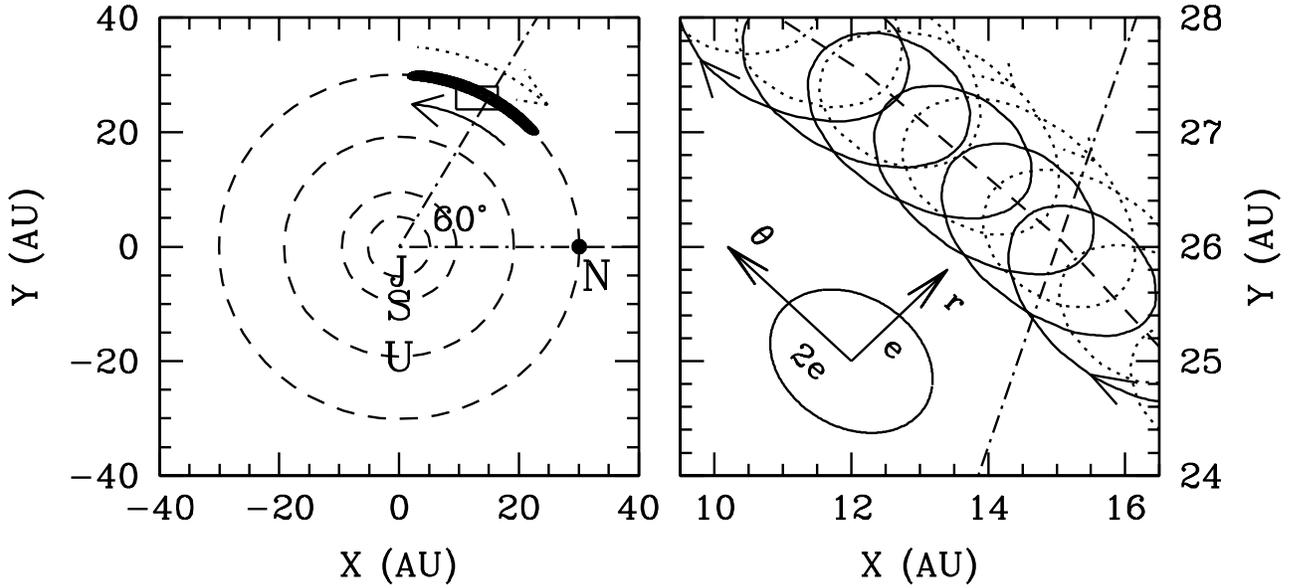}
\vspace{-2in}
\caption{Trajectory of the Neptune Trojan, 2001 QR$_{322}$ (``QR''),
numerically integrated in the presence of the giant planets
(denoted J, S, U, N) for $10^4 \yr$ and viewed from
above the plane of the Solar System.
In the left-hand panel, the tube
of densely packed points traces QR's trajectory; the Sun sits
at the origin, the distance of each point
from the origin equals QR's instantaneous
heliocentric distance, and the angle
that the Sun-QR vector makes with respect to the X-axis
equals the instantaneous angle between the Sun-QR and
Sun-Neptune vectors.
The inset box is magnified in the right-hand panel, which shows
individual epicycles and their relative dimensions in the radial
and azimuthal directions. Each epicycle completes in $t_{\rm orb} =
2\pi/\Omega \approx 160\yr$, while the guiding center of the
epicycle loops around L4 every $t_{\rm lib} \approx 8.9 \times 10^3 \yr$.
Arrows in both panels indicate directions of motion.
}
\label{f1}
\end{figure}

Figure \ref{f1} illustrates QR's trajectory: a combination
of epicyclic and guiding center motion in the frame
co-rotating with the planet. The Trojan
completes each ellipse-shaped epicycle (``corkscrew turn'')
in an orbit time,

\begin{equation}
t_{\rm orb} = \frac{2\pi}{\Omega} \approx \frac{2\pi}{\Omega_{\rm p}} \approx
160 \yr \,,
\end{equation}

\ni where $\Omega$ and $\Omega_{\rm p}$ are the object's and the host
planet's orbital angular frequencies, respectively. Projected
onto the host planet's orbit plane, the epicycle's
major and minor axes
align with the azimuthal and radial directions, respectively.
The ratio of semi-axis lengths is $2ea$ : $ea$,
where $e$ and $a \approx a_{\rm p}$ are, respectively,
the osculating eccentricity and orbital semi-major axis
of the Trojan, and $a_{\rm p}$ is the orbital semi-major axis
of the host planet.

\placefigure{fig1}

\subsection{Guiding Center Motion}
\label{gcmotion}
The guiding center of the epicycle loops around the Lagrange
point every libration period,

\begin{equation}
t_{\rm lib} \approx \left( \frac{4}{27\mu} \right)^{1/2}\, t_{\rm orb} \approx
8.9 \times 10^{3}\yr \,,
\label{alib}
\end{equation}

\ni where $\mu = M_{\rm N}/M_{\odot} \approx 5 \times 10^{-5}$ is
the ratio of Neptune's mass to the Sun's. This analytic expression
for $t_{\rm lib}$ is given by linear stability analysis
and is independent of the size of the guiding center orbit.
We supply a more precise formula that depends on orbit size
in \S\ref{anharm}. The guiding center traces approximately
an extremely elongated ellipse (``tadpole'') centered on the Lagrange
point and having a radial : azimuthal aspect ratio of $(3\mu)^{1/2}$ : $1$,
as depicted in Figure \ref{nordicell}.
The semi-minor axis of the largest possible tadpole has a length of

\begin{equation}
\max(\delta a) \approx \left( \frac{8\mu}{3} \right)^{1/2} a_{\rm p} \,.
\end{equation}

\ni This length equals
the greatest possible difference between the osculating
semi-major axes of the Trojan and of the host planet.

\subsection{Trojan-Trojan Encounters}
\label{trojtroj}
\begin{figure}
\epsscale{0.9}
\vspace{-1in}
\plotone{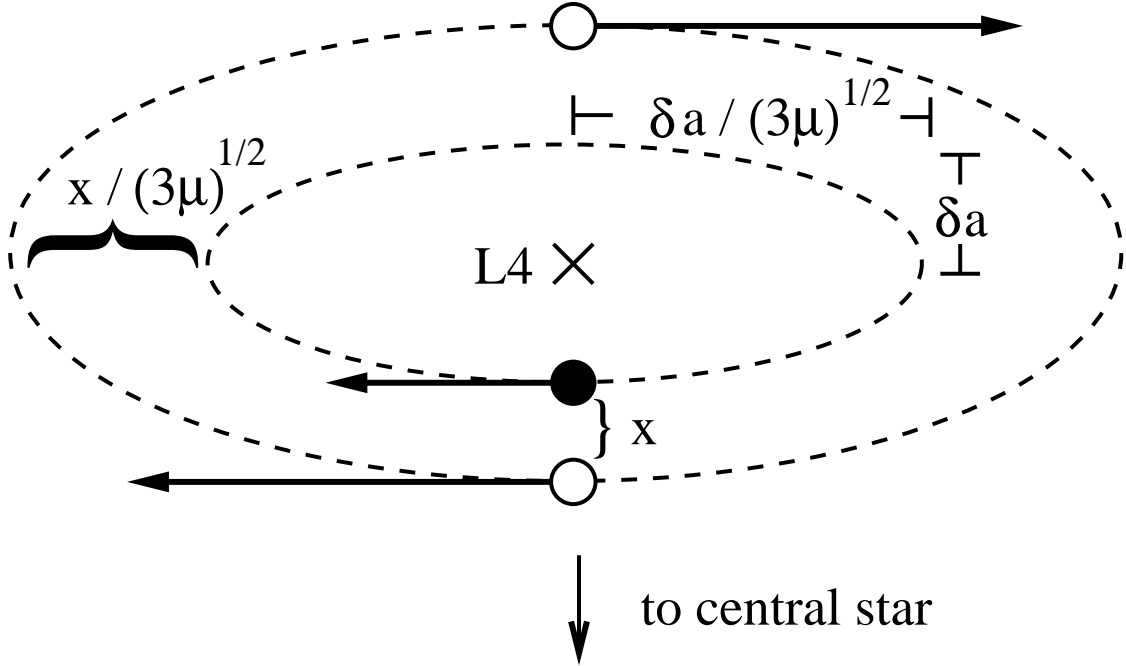}
\caption{Relative motions of orbital guiding centers in Trojan resonance.
Each guiding center executes an elliptical trajectory (``tadpole'')
around L4 having a radial : azimuthal
aspect ratio of $(3\mu)^{1/2}$:1. One Trojan (bottom open circle)
is shown undergoing a ``close'' conjunction with another (solid circle).
The relative velocity of guiding centers
during a close conjunction is given approximately by
Keplerian shear. Because bodies are in Trojan resonance,
they execute one loop around L4 every $t_{\rm lib}$.
Were it not for anharmonic shear, close conjunctions
between the two bodies would occur every $t_{\rm syn} =
t_{\rm lib}/2$, alternately
above and below L4. Given anharmonic shear, any given pair of Trojans
eventually undergoes ``distant'' conjunctions, of which
one is also shown (top open and solid circles).
}
\label{nordicell}
\end{figure}

Consider two Trojans moving initially on pure tadpole orbits with
zero epicyclic amplitudes (Figure \ref{nordicell}, bottom solid and open
circles).
A ``close'' conjunction between the bodies
occurs with radial separation $x$, at a location
away from the turning points of either tadpole orbit and at a time
when both bodies are moving in the same direction.\footnote{Throughout
this paper, we use the word ``conjunction'' in the usual heliocentric
sense; two bodies undergoing a conjunction are collinear
with the Sun, and not necessarily with the L4 point.}
The dynamics during the conjunction are, to good approximation, like those of
a conjunction between two bodies on circular Keplerian
orbits in the absence of the planet. That is, the relative velocity
is $\sim$$3\Omega x / 2$ and therefore the duration of the
encounter is $\sim$$1/\Omega$. We have verified that this is the
case by numerical orbit integrations. 

Close conjunctions occur
twice per libration period, radially inside and outside the L4 point.
We define a synodic time,

\begin{equation}
t_{\rm syn} \equiv t_{\rm lib}/2 \,,
\end{equation}

\ni between successive close conjunctions.
Two bodies separated by a radial distance $x$
and an azimuthal distance $\lesssim L_{\rm stir}(x) \equiv 3\Omega x t_{\rm
syn}/2$ undergo close conjunctions every $t_{\rm syn}$.

\placefigure{fig2}

\subsection{Anharmonic Shear}
\label{anharm}
If the libration period, $t_{\rm lib}$, were truly independent
of tadpole size, two bodies undergoing
close conjunctions would undergo them indefinitely; two conjunctions
would occur at the same locations
with respect to L4 every libration period.\footnote{Under such a supposition
it would nonetheless be incorrect to say that the Trojan libration
region is in solid body rotation, since close conjunctions
still involve Keplerian shear, as discussed in \S\ref{trojtroj}.}
In fact, anharmonicity
of the perturbation potential causes the libration period
to grow with tadpole size. We calculate numerically
the deviation,

\begin{equation}
\delta t_{\rm lib} \equiv t_{\rm lib} - t_{\rm  lib,0} \,,
\end{equation}

\begin{figure}
\epsscale{0.9}
\plotone{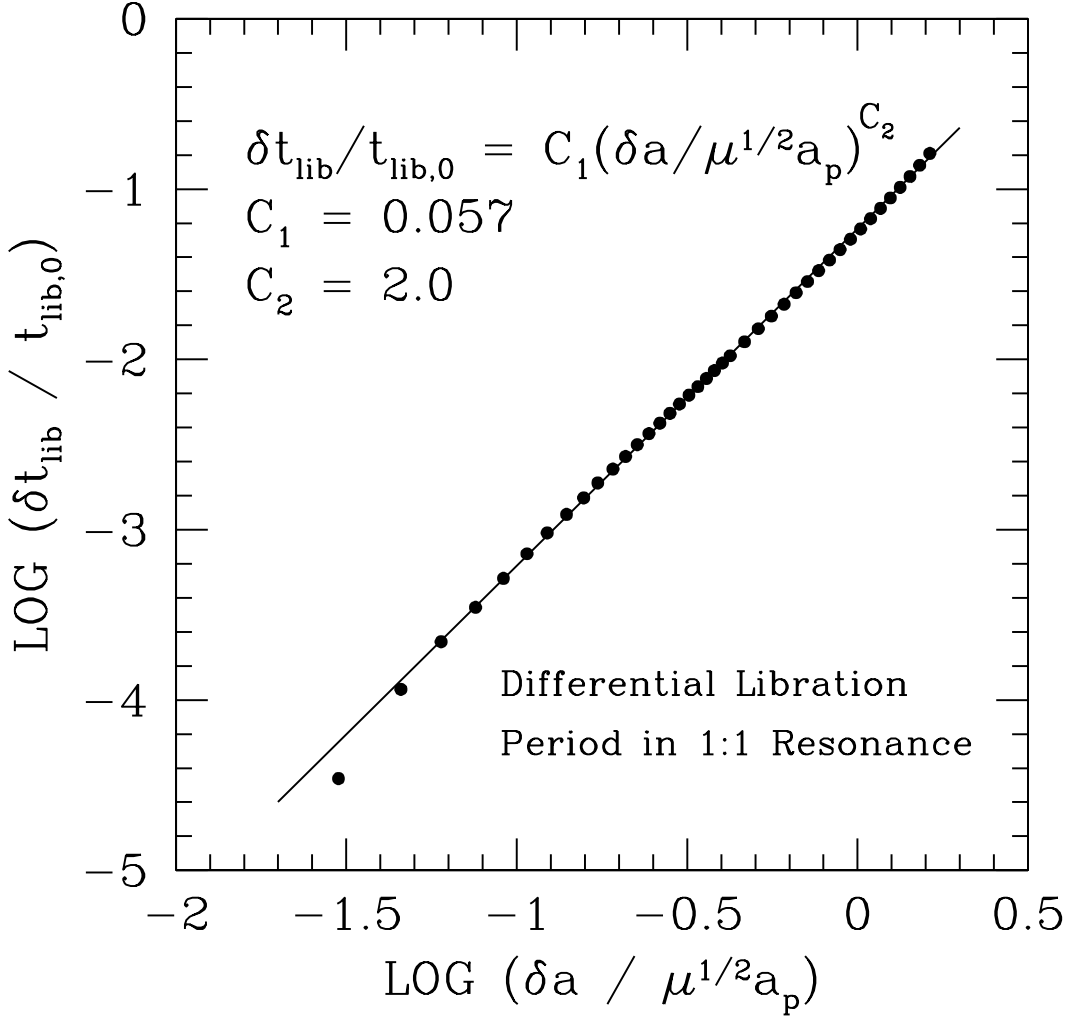}
\caption{Increase of libration period with tadpole size, as measured
by numerical integrations of pure tadpole orbits for $\mu = 5 \times 10^{-5}$
in the circular, restricted, planar 3-body problem. Tadpole size
is described by $\delta a$, the tadpole
semi-minor axis (see Figure \ref{nordicell}). As $\delta a$
increases, the difference, $\delta t_{\rm lib}$,
between the measured libration period, $t_{\rm lib}$,
and a fiducial libration period, $t_{\rm lib,0}$
(evaluated at $\delta a = 10^{-4} a_{\rm p}$),
grows. The difference is fitted to a power law (solid line) having
the parameters displayed. The increase of libration
period with tadpole size (``anharmonic shear'')
causes close conjunctions to give way to distant
conjunctions over the grand synodic time,
$t'_{\rm syn}$.
}
\label{tadshear}
\end{figure}

\ni as a function of tadpole semi-minor axis, $\delta a$,
where $t_{\rm lib,0}$ is the libration period for $\delta a = 10^{-4}
a_{\rm p}$.
We integrate, via the Burlirsch-Stoer algorithm (Press et al.~1992),
test particle orbits having virtually zero epicyclic amplitudes
($ea \ll \delta a$)
in the presence of a binary of mass ratio $\mu = 5 \times 10^{-5}$
and orbital eccentricity $e_{\rm p} = 0$.
The numerically computed value of $t_{\rm lib,0}$
matches that calculated from our
analytic expression (\ref{alib}) to within 1 part in $10^5$.
The result for $\delta t_{\rm lib}$, documented in Figure \ref{tadshear},
may be fitted to a power law,

\begin{equation}
\delta t_{\rm lib} = 0.057 \, t_{\rm lib,0} \left( \frac{\delta
  a}{\mu^{1/2}a_{\rm p}} \right)^{2.0} \,.
\end{equation}

\ni The anharmonicity is never strong; $\delta t_{\rm lib} < t_{\rm lib}$.

\placefigure{fig3}

This ``anharmonic shear'' permits phase differences to accumulate
between neighboring Trojans. Close conjunctions occur
for a finite number of libration periods; eventually, close
conjunctions give way to ``distant'' conjunctions
that occur when both bodies are moving in opposite directions
on radially opposite sides of L4 (Figure \ref{nordicell}, top open and solid
circles). Each cycle of close-distant-close conjunctions lasts for a time,

\begin{equation}
t'_{\rm syn} \approx  \left| \frac{d (t_{\rm lib}^{-1})}{d (\delta a)} x
\right|^{-1} \approx 8.8 \, t_{\rm lib,0} \left( \frac{\mu^{1/2}a_{\rm
p}}{\delta a} \right)^{2.0} \frac{\delta a}{x} \,,
\end{equation}

\ni where $x \ll \delta a$ is the minimum (radial) distance between neighboring
tadpole
orbits. For a typical value of $\delta a = 0.5 \mu^{1/2} a_{\rm p}$,
this ``grand synodic time'' takes a simple form:

\begin{equation}
t'_{\rm syn} \approx 43 \frac{a_{\rm p}}{\Omega_{\rm p} x} \,,
\label{gst}
\end{equation}

\ni which is similar to, but of order 10 times longer than,
the ordinary synodic time in a circular Keplerian disk
away from resonance ($4\pi a_{\rm p}/3\Omega_{\rm p} x$).

The time spent undergoing close conjunctions, $t_{\rm stir}$,
is shorter than $t'_{\rm syn}$
by the ratio of $L_{\rm stir}$ to the circumference
of a Trojan's guiding center orbit:

\begin{equation}
t_{\rm stir} \approx \frac{\pi}{4} \frac{x}{\delta a} \, t'_{\rm syn} \approx
56 \, t_{\rm syn} \,,
\end{equation}

\ni independent of $x$.
The number of conjunctions that occur during this ``stirred'' phase is 
$t_{\rm stir}/t_{\rm syn}\approx 56$.




\subsection{Summary}
\label{summ}
To summarize the behavior described in sections
\S\S\ref{epimotion}--\ref{anharm}: For a time
$t_{\rm stir}$, two neighboring Trojans whose underlying
tadpole orbits have semi-minor axes (measured relative to L4)
of $\delta a$ and $\delta a + x$ undergo close
conjunctions. These conjunctions occur every $t_{\rm syn} \approx
t_{\rm lib}/2 \approx 4.5 \times 10^3 \yr$,
and are like ordinary conjunctions between bodies
on circular orbits outside of resonance. In particular,
an individual conjunction, during
which the distance between guiding centers is $\sim$$x$,
lasts $\sim$$1/\Omega$. The number of conjunctions that occur during this
stirred phase is typically $t_{\rm stir}/t_{\rm syn} \sim 56$.
After $t_{\rm stir}$ time elapses,
sufficient phase accumulates between the Trojans that
close conjunctions cease. 
The ``unstirred'' phase, during which distant conjunctions occur
and the distance between bodies is $\gg x$,
lasts $t'_{\rm syn} \sim (\delta a / x) t_{\rm stir}$.
Afterwards, close conjunctions resume.

\section{PROPERTIES OF THE NEPTUNE TROJAN POPULATION}
\label{character}
We review the properties of Neptune Trojans requiring explanation.
With only one Trojan known, we infer many of these characteristics
by combining direct observations with theory.

\subsection{Orbit}
\label{orbit}
Evaluated in a heliocentric, J2000 ecliptic-based coordinate system
on Julian date 2451545.0,
the osculating elements of QR are $a = 30.1$ AU,
$e = 0.03$, and $i = 0.02$ rad (Elliot et al.~2005).
Uncertainties in these values are less than 10\% ($1\sigma$)
and computed according to the procedure developed
by Bernstein \& Khushalani (2000).
The epicycles traced by QR are larger than the radial width
of the tadpole; $e,i > \mu^{1/2} = 0.007$.

The libration amplitude,

\begin{equation}
\Delta\phi = \max(\lambda-\lambda_{\rm p}) - \min(\lambda-\lambda_{\rm p})\,,
\end{equation}

\ni equals the full angular extent of the tadpole orbit,
where $\lambda$ and $\lambda_{\rm p}$ are the mean
orbital longitudes of the Trojan and of the planet, respectively.
For QR, $\Delta\phi \approx 48^{\circ}$ (C03).

\subsection{Physical Size}
\label{size}
An assumed albedo of 12--4\% yields a radius for QR
of $R \sim 65$--$115 \km$ (C03).
This size is comparable to that of the largest
known Jupiter Trojan, (624) Hektor,
whose minimum and maximum semi-axis lengths
are $\sim$75 km and $\sim$150 km, respectively
(Barucci et al.~2002).\footnote{Hektor might be a near-contact binary (Hartmann
\& Cruikshank 1978; Tanga et al.~2003).} We refer to Trojans resembling QR as
``large.'' 

\subsection{Current Number}
\label{currpop}
The distribution of DES search fields on the sky, coupled with theoretical
maps of the sky density of Neptune Trojans (Nesvorny \& Dones 2002),
indicate that $N \sim 10$--30 large objects (resembling QR) librate
about Neptune's L4 point (C03).

If the true radius of QR is near our estimated lower bound, $R \sim 65$ km,
then the number of large Neptune Trojans is comparable
to that of large Jupiter Trojans, of which there are
$\sim$10 whose radii exceed $\sim$65 km (Barucci et al.~2002).
If the true radius of QR is closer to our estimated upper bound,
$R \sim 115$ km, then large Neptune Trojans
outnumber their Jovian counterparts by
a factor of $\sim$10--30, since there is only
1 Jupiter Trojan (Hektor) whose radius exceeds $\sim$100 km
(Barucci et al.~2002).

\subsection{Past Number: Collisional Attrition}
\label{pastpop}
Large Neptune Trojans observed today are essentially
collisionless; they are not
the remains of a once greater population
that has been reduced in number by collisions.
We consider here catastrophic
dispersal by collisions with bodies in the
same Trojan cloud. By catastrophic dispersal we mean
that the mass of the largest post-collision fragment is no greater
than half the mass of the original target and that collision
fragments disperse without gravitational re-assembly.
The lifetimes, $t_{\rm life}$, of large Neptune Trojans
against catastrophic dispersal
depend on their relative velocities, $v_{\rm rel}$, at impact.
If the orbit of QR is typical, then
$v_{\rm rel} \sim \Omega a\sqrt{e^2 + i^2} \sim 200 \m/\rm{s}$.
At such impact velocities, catastrophically
dispersing a target of radius $R \sim 90 \km$
and corresponding mass $M$ may be impossible. This is seen as follows.
The gravitational binding energy of such a target
well exceeds its chemical cohesive energy.
Then dispersal requires a projectile mass $m$ satisfying

\begin{equation}
\frac{f_{\rm KE}}{2} \left( \frac{mM}{m+M} \right) v_{\rm rel}^2 \, \gtrsim
\, \frac{3G(M+m)^2}{5R_{M+m}} \,,
\label{breakup}
\end{equation}

\ni where $R_{M+m}$ is the radius of the combined mass $M+m$,
and  $f_{\rm KE}$
is the fraction of pre-collision translational kinetic energy
converted to post-collision translational kinetic energy
(evaluated in the center-of-mass frame of $m$ and $M$).
Observed properties of Main Belt asteroid
families and laboratory impact experiments
suggest $f_{\rm KE} \sim 0.01$--0.1 (Holsapple et al.~2002;
Davis et al.~2002).
Equation (\ref{breakup}) may be re-written

\begin{equation}
\frac{m/M}{(1+m/M)^{8/3}} \gtrsim \frac{8\pi G\rho R^2}{5 f_{\rm KE} v_{\rm
rel}^2} \,,
\label{breakup2}
\end{equation}

\ni
where $\rho \sim 2 \gm \cm^{-3}$ is the internal mass density of an object.
For the parameter values cited above, the right-hand-side of equation
(\ref{breakup2}) equals $1.4 \, (0.1/f_{\rm KE})$.
Since the maximum of the left-hand-side
is only 0.17, catastrophic dispersal cannot occur at such low
relative velocities.


Even if the inclination dispersion of Neptune Trojans were instead
as large as $\sim$0.5 rad---similar to that of Jupiter Trojans, and permitted
by dynamical stability studies (Nesvorny \& Dones 2002)---collisional
lifetimes are probably too long to be significant. Such a large
inclination dispersion would imply that
$v_{\rm rel} \sim 2.7 \km\,\s^{-1}$ and that 
projectiles having radii $r \sim 20 \km$ could
catastrophically disperse QR-sized objects [by equation (\ref{breakup2})].
The number, $N_{r}$, of such projectiles is unknown.
If the size distribution of Neptune Trojans resembles
that of Jupiter Trojans (Jewitt et al.~2000),
then $N_{r} \sim 6000$, which would imply

\begin{equation}
t_{\rm life} \sim \frac{4}{\pi} \frac{e_r i_r}{N_{r}\sqrt{e_r^2 + i_r^2}}
\left( \frac{a}{R} \right)^2 \frac{1}{\Omega} \sim 4 \times 10^{11} \left(
\frac{6000}{N_{r}} \right) \left( \frac{90\km}{R} \right)^2 \yr \gg t_{\rm sol}
\, ,
\label{tlife}
\end{equation}

\ni where we have taken Trojan projectiles to occupy
a volume of azimuthal arclength $\sim$$a$,
radial width $\sim$$2e_r a$, and vertical height $\sim$$2i_r a$,
and inserted $e_r \approx 0.03$ and $i_r \approx 0.5$.




We conclude that the current number of QR-sized
Neptune Trojans cannot be explained by appealing to destructive
intra-cloud collisions.
That large Neptune Trojans have not suffered collisional attrition
suggests that their number directly reflects the efficiency of a primordial
formation mechanism.

\subsection{Past Number: Gravitational Attrition}
\label{graver}

\subsubsection{Attrition During the Present Epoch}
It seems unlikely that gravitational
perturbations exerted by Solar System planets in their
current configuration reduced the Neptune Trojan
population by orders of magnitude. Nesvorny \& Dones (2002)
perform the following experiment that suggests
Neptune Trojans are generically stable in the present epoch. They 
synthesize a hypothetical population of 1000 Neptune Trojans
by scaling the positions and velocities of actual Jupiter Trojans.
Over $t_{\rm sol}$, about 50\% of their Neptune
Trojans remain in resonance. Objects survive
even at high inclination, $i \sim 0.5$.

\subsubsection{Today's Trojans Post-Date Neptune's Final Circularization}
\label{postdate}
By contrast, during the era of planet formation,
dramatic re-shaping of the planets' orbits
likely impacted the number of Neptune Trojans significantly.
To pinpoint the time of birth of present-day Trojans,
we must understand the history of Neptune's orbit.
Current conceptions of this history
involve a phase when proto-Neptune's orbit was strongly perturbed
by neighboring protoplanets (GLS04;
see also Thommes et al.~1999).\footnote{Our discussion assumes that the
ice giant cores formed beyond a heliocentric distance of $\sim$20 AU
by accreting in the sub-Hill regime (GLS04).
Our conclusion that present-day Neptune
Trojans formed after Neptune's orbit finally circularized does not
change if we follow instead the scenario in which the ice giant cores were
formed between Jupiter and Saturn
and were ejected outwards (Thommes et al.~1999).
}
Once protoplanets grew to when their mass became
comparable to the mass in remaining planetesimals,
circularization of the protoplanets' orbits by dynamical friction
with planetesimals was rendered ineffective (GLS04; see also \S\ref{insitu}).
Subsequently, the protoplanets gravitationally scattered
themselves onto orbits having eccentricities of order unity.
At this time, the mass
of an individual protoplanet equaled the isolation mass,

\begin{equation}
M_{\rm iso} \sim (4\pi A\sigma)^{3/2} a_{\rm p}^3 (3M_{\odot})^{-1/2} \gtrsim 3
M_{\oplus} \, ,
\label{miso}
\end{equation}

\ni which is the mass enclosed within each protoplanet's annular
feeding zone of radial width
$\sim$$2AR_{\rm H,p}$, where
$R_{\rm H,p} = (M_{\rm iso}/3M_{\odot})^{1/3}a_{\rm p}$
is a protoplanet's Hill sphere radius,
$A \approx 2.5$ (Greenberg et al.~1991), and $M_{\oplus}$
is an Earth mass. For the surface
density, $\sigma$, of the protoplanetary disk,
we use $\sigma \gtrsim \sigma_{\rm min} \sim 0.2 \g/{\rm cm}^2$,
where $\sigma_{\rm min}$ is
the surface density of condensates in the minimum-mass solar
nebula at a heliocentric distance of 30 AU.
The actual surface density might well have exceeded
the minimum-mass value by a factor of $\sim$3, in which case
$M_{\rm iso} = M_{\rm N} = 17 M_{\oplus}$.
The phase of high eccentricity ended when enough protoplanets were ejected
from the Solar System that proto-Neptune's orbit could once again
be kept circular by dynamical friction with planetesimals.

\begin{figure}
\epsscale{0.8}
\plotone{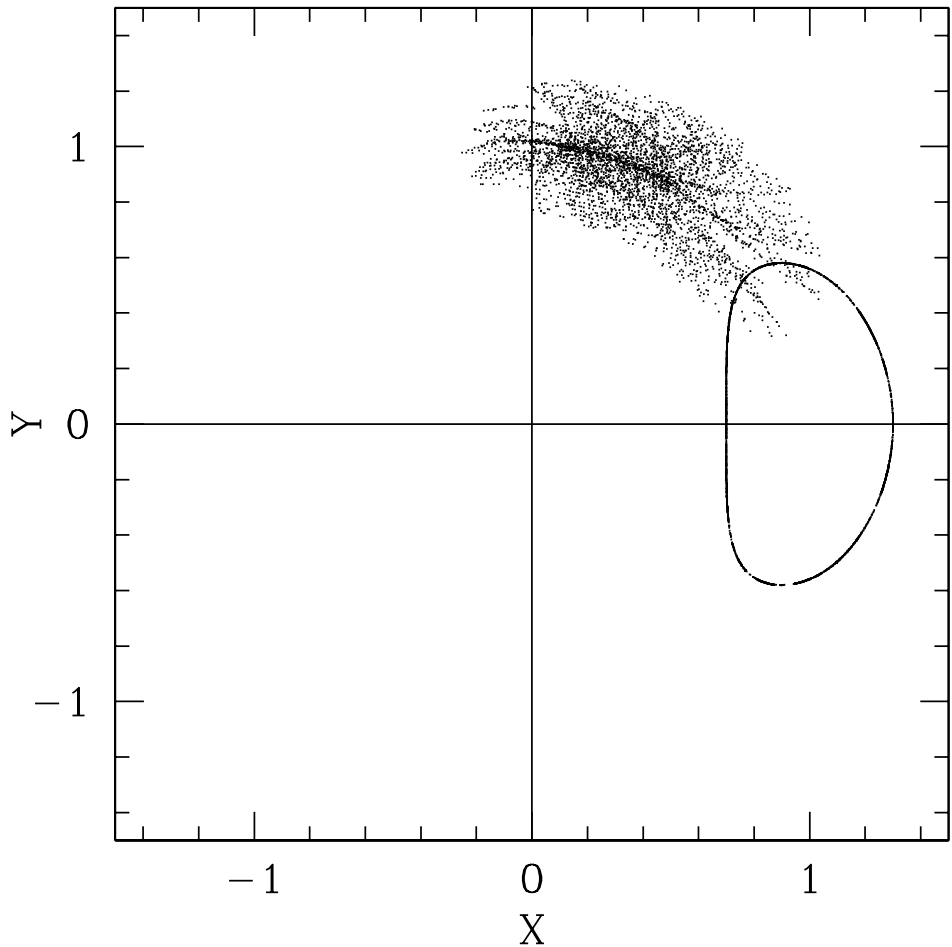}
\caption{Trajectory of a Trojan test particle
hosted by a planet ($\mu = 5\times 10^{-5}$)
moving on an orbit of eccentricity $e_{\rm p} = 0.3$. Positions $X$
and $Y$ are in units of the semi-major axis of the planet-star binary,
and are measured relative to the central star
in the frame rotating at the binary mean motion.
The planet was initially located at periastron along the
$X$-axis. Initial conditions for the test particle were such that if
$e_{\rm p} = 0$, the test particle would be nearly stationary at L4.
Scattered points indicate positions of the Trojan sampled over 1000 orbital
periods, while the near-solid curve traces the epicyclic
trajectory of the planet.
The tadpole region occupied by the Trojan is as radially wide as
the planet's epicycle.
Trojan orbits would have crossed those
of nearby protoplanets during Neptune's high eccentricity phase;
formation of present-day Neptune Trojans must wait until after
Neptune's orbit finally circularized.}
\label{ecchost}
\end{figure}

Trojans present prior to Neptune's high-eccentricity
phase would likely have escaped the resonance due to perturbations
by neighboring protoplanets. Our numerical integrations show that while
Trojans can be hosted by highly eccentric planets, such Trojans undergo
fractional excursions in orbital radius
as large as those of their hosts, i.e., of order unity (Figure \ref{ecchost}).
Since fractional separations between protoplanets
would only have been of order $2AR_{\rm H,p}/a_{\rm p} \sim 0.1$,
Trojan orbits would have crossed those of nearby protoplanets.
Long-term occupancy of the resonance must wait until after
protoplanet ejection and the final circularization of Neptune's orbit.

\placefigure{fig4}

Note that C03 argue that the existence of Neptune Trojans
rules out violent orbital histories for Neptune. We consider
their case to be overstated. Present-day Neptune Trojans
can be reconciled with prior eccentricities of order unity
for Neptune's orbit, provided that Trojan formation occurs
after Neptune's circularization by dynamical friction.

\subsubsection{Attrition During the Epoch of Migration}
\label{eromig}

After Neptune's orbit circularized, the planet may still
have migrated radially outwards by scattering
ambient planetesimals (Fernandez \& Ip 1984;
Hahn \& Malhotra 1999).
Neptune Trojans can escape during
migration by passing through secondary
resonances with Uranus and the gas giants (Gomes 1998;
Kortenkamp, Malhotra, \& Michtchenko~2003).
In our analysis below, we assume that Trojans form
after migration and therefore do not suffer such losses.

\section{FORMATION BY PULL-DOWN CAPTURE}
\label{pulldown}
Trojans can, in principle, be captured via mass growth of the host planet.
This mechanism, which we call ``pull-down capture,''
has been proposed to generate Jupiter Trojans as a massive gaseous envelope
accreted onto the core of proto-Jupiter (Marzari \& Scholl 1998ab;
Fleming \& Hamilton 2000; Marzari et al.~2002).\footnote{``Pull-down capture''
was coined to describe capture of bodies onto planetocentric
(satellite) orbits by mass growth of the planet (Heppenheimer \& Porco 1977).}

Pull-down capture may have played a supporting role
in the capture of Neptune Trojans, but likely not a leading one.
The mechanism operates on the principle of adiabatic invariance.
If, as is likely, the mass of the planet grows on a timescale longer than
$t_{\rm lib}$, then the libration amplitudes of 1:1 resonators
shrink as

\begin{equation}
\Delta\phi \propto \mu^{-1/4} \,.
\label{weakling}
\end{equation}

\ni Horseshoe orbits---those in 1:1 resonance that loop around both 
triangular points, so that $\Delta\phi \gtrsim 320^{\circ}$---can be
converted to Trojan orbits, having
$\Delta \phi \lesssim 160^{\circ}$. These
bounds derive from the circular, restricted, planar 3-body problem.
One shortcoming of current treatments of pull-down capture is that
the prior existence of horseshoe librators is assumed
without explanation. Horseshoe librators are
more unstable than tadpole librators; the former
escape resonance more easily due to perturbations
by neighboring planets.\footnote{Karlsson (2004)
has identified $\sim$20 known asteroids that, sometime
within 1000 years of the current epoch, occupy
horseshoe-like orbits with Jupiter. These objects transition often
between resonant and non-resonant motion.}
What sets the number of these
weakly bound resonators at the beginning of pull-down scenarios
is unclear.

Even if we ignore the problem
of having to explain the origin of horseshoe librators,
the efficacy of pull-down capture is weak [equation (\ref{weakling})].
The factor by which Neptune
increases its mass subsequent to its high-eccentricity phase is
$M_{\rm N}/M_{\rm iso} \lesssim 6$ [equation (\ref{miso})].
Such mass growth implies that pull-down capture, operating alone,
produces Trojans having only large libration
amplitudes, $160^{\circ} \gtrsim \Delta\phi \gtrsim 160^{\circ}/6^{1/4} \sim
100^{\circ}$.
By contrast, the orbit of QR is characterized by
$\Delta\phi \approx 48^{\circ}$. Excessive libration amplitudes
even afflict Jupiter Trojans formed
by pull-down capture, despite the $\sim$30-fold
increase in Jupiter's mass due to gas accretion (Marzari \& Scholl 1998b).
Collisions have been proposed to extend the distribution
of libration amplitudes down to smaller values (Marzari \& Scholl
1998b), but our analysis in \S\ref{pastpop} indicates that
large Trojans are collisionless. Moreover, inter-Trojan
collisions dissipate energy and therefore
deplete the resonance; see \S\ref{intro} and \S\ref{insitu}.

In sum, formation of large Neptune Trojans by pull-down
capture seems unlikely because libration amplitudes
are inadequately damped; Neptune increases
its mass by too modest an amount after the planet's
high-eccentricity phase.

\section{FORMATION BY DIRECT COLLISIONAL EMPLACEMENT}
\label{deposit}

Large, initially non-resonant objects of radius $R \sim 90 \km$
can be deposited directly into Trojan resonance by collisions.
Successful deflection of target mass $M$
by projectile mass $m$ requires that the post-collision semi-major
axis of $M$ lie within $\sim$$\mu^{1/2}a_{\rm p}$ of $a_{\rm p}$,
and that the post-collision eccentricity be small (resembling that of QR).
We estimate the number of successful depositions into one Trojan cloud as
follows. First, we recognize that successful emplacement
requires $m \sim M$, since it is difficult for widely varying
masses to significantly deflect each other's trajectory. This
will be justified quantitatively in \S\ref{fcolmethod}.
Each target of mass $M$ and radius $R$ collides with

\begin{equation}
U_{\rm col} \sim n_M R^2 v_M
\end{equation}

\ni similar bodies per unit time, where

\begin{equation}
n_M \sim \frac{\Sigma_M \Omega_{\rm p}}{M v_M}
\end{equation}

\ni is the number density of bodies, $\Sigma_M$
is their surface density, and $v_M$ is their
velocity dispersion (assumed isotropic).
Of the collisions occurring within a heliocentric annulus
centered at $a = a_{\rm p}$ and of radial width $\Delta a = a_{\rm p}/2$,
only a fraction, $f_{\rm col}$,
successfully deflect targets onto low eccentricity
orbits around one Lagrange point.
After time $t_{\rm col}$ elapses, the number of successful emplacements
into one cloud equals

\begin{eqnarray}
N_{\rm col} & \sim & \frac{2\pi\Sigma_M a_{\rm p} \Delta a}{M}\, U_{\rm col}
f_{\rm col} t_{\rm col} \label{coll}\\
& \sim & \frac{2\pi \Omega_{\rm p} a_{\rm p} \Delta a R^2}{M^2} \,\Sigma_M^2
f_{\rm col} t_{\rm col} \,.
\label{coll2}
\end{eqnarray}

\ni In \S\ref{fcolmethod}, we detail our method of computing
$f_{\rm col}$. We describe and explain the
results of our computations in \S\ref{fcolres}.
Readers interested only in the 
consequent demands on $\Sigma_M$ and $t_{\rm col}$
and whether they might be satisfied may skip
to \S\ref{upshot}.

\subsection{Method of Computing $f_{\rm col}$}
\label{fcolmethod}

For fixed target mass $M$ and projectile mass $m$,

\begin{equation}
f_{\rm col} = \frac{1}{\Delta a} \int_{a_{\rm p} - \Delta a/2}^{a_{\rm
p}+\Delta a/2} f \, da_M \,,
\end{equation}

\ni where $a_M$ is the pre-collision
semi-major axis of mass $M$,
and $f$ is the probability that a collision
geometry drawn randomly from the distribution of pre-collision
orbits yields a successfully emplaced Trojan. We provide a more precise
definition of success below.
For the collision geometries that we experiment
with, we find that $\Delta a = 0.5 a_{\rm p}$ ensures
that all successful collisions are counted (i.e., $f$ goes
to zero at the limits of integration).

Computing $f_{\rm col}$ requires knowing how pre-collision
semi-major axes, eccentricities, and inclinations are distributed. 
Since these distributions in the early Solar System are unknown,
we attempt the more practicable
goal of estimating the maximum value of $f_{\rm col}$
by experimenting with simple cases. To better understand the
ingredients for a successful emplacement, we allow $m \neq M$.
We model collisions as completely inelastic encounters
between point particles, though we allow for the possibility of
catastrophic dispersal. These simplifications 
permit maximum deflection of $M$'s trajectory and imply that
the outcome of a non-catastrophic collision is a merged body of mass $M+m$.

We adopt distributions of pre-collision orbital elements
as follows. 
Both $M$ and $m$ are taken to have the same distribution
of orbital guiding centers (semi-major axes)
located outside the planet's Hill sphere. Given input parameter $B \geq 1$,
semi-major axes of masses $M$ and $m$ are uniformly distributed
over values greater than $(1+B\mu^{1/3})a_{\rm p}$
and less than $(1-B\mu^{1/3})a_{\rm p}$, but take no intermediate
value. The error incurred in writing equation (\ref{coll})
without regard to the evacuated Hill sphere region
is small for $B\mu^{1/3} < \Delta a / a_{\rm p}$.
Eccentricities of masses $M$ are fixed at $e_M = C \mu^{1/3}$,
and those of masses $m$ are fixed at $e_m = D \mu^{1/3}$,
where constants $C, D \gtrsim B$ to ensure that
bodies wander into the Trojan libration zone.
Finally, target and projectile are assumed to occupy
co-planar orbits. The condition of co-planarity
is relaxed in \S\ref{fcolres}; for now, we note that
allowing for mutual inclination increases the relative velocity
at impact and tends to produce catastrophic dispersal
and large post-collision eccentricities, reducing $f_{\rm col}$.

Calculations are performed for fixed $M$ appropriate
to $R = 90\km$ and $\rho = 2\gm\cm^{-3}$.
For each $B$, $C$, $D$, $m$, $a_{M}$, and
true anomaly (angular position from pericenter)
of mass $M$, all possible orbits of $m$ that collide with $M$
are computed. This set of possible orbits is
labelled by the true anomaly of $m$
at the time of collision.
Post-collision semi-major axes and eccentricities are
computed by conserving momentum in the radial and azimuthal
directions separately.
Successful emplacements involve (a)
post-collision semi-major axes 
of the merged body that lie within
$\pm \mu^{1/2}a_{\rm p}$ of $a_{\rm p}$;
(b) no catastrophic
dispersal, where the criterion for dispersal
is given by equation (\ref{breakup}) and two values
of $f_{\rm KE}$ are tested, 0.01 and 0.1;
and (c) post-collision eccentricities
less than 0.05. This last
requirement is motivated by QR's eccentricity ($e = 0.03$)
and the tendency for more eccentric objects
to be rendered unstable by Uranus.
Successful collisions are tallied over all true
anomalies of $m$ and $M$,
divided by the total number of collision geometries
possible, and divided further by $2\pi$ to yield $f$.
The last division by $2\pi$ accounts for the probability
that the collision occurs at the appropriate azimuth
relative to Neptune, in an arc of
azimuthal extent $\sim$1 rad centered on L4.
For the small-to-moderate eccentricities considered
here, every interval in true anomaly is taken to occur with
equal probability.

\subsection{Results for $f_{\rm col}$}
\label{fcolres}

\begin{figure}
\epsscale{0.8}
\plotone{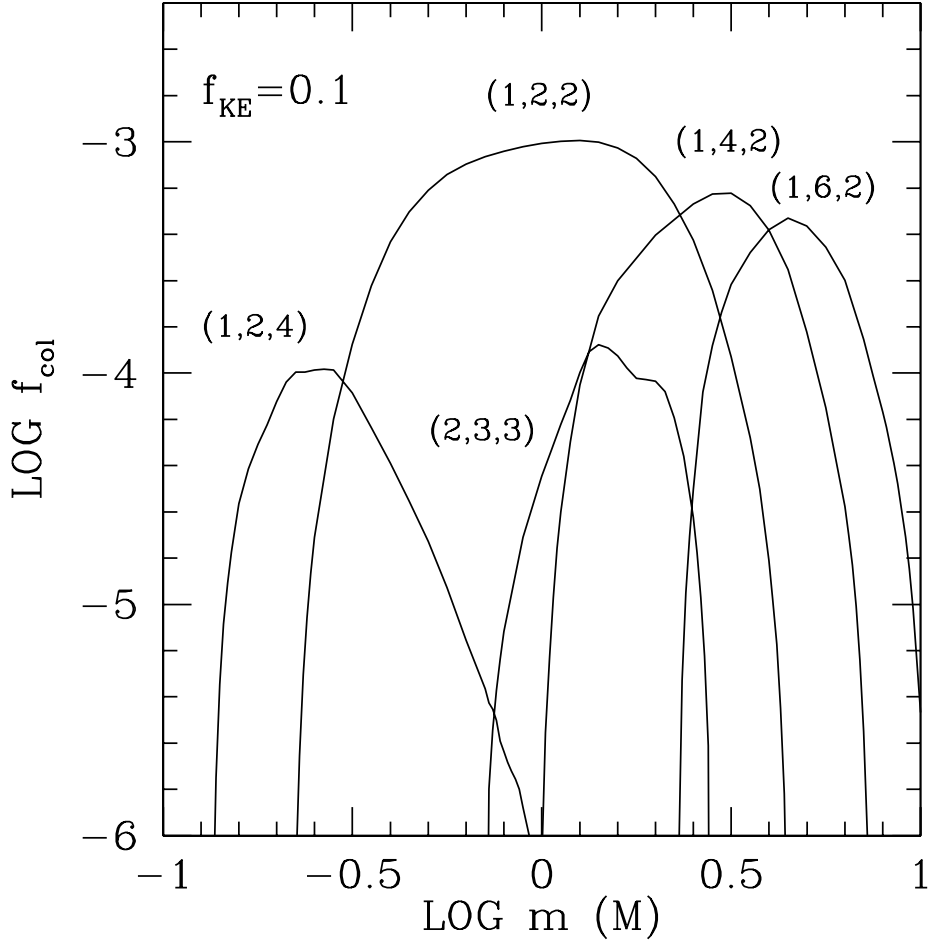}
\caption{Efficiency factors, $f_{\rm col}$, for direct emplacement
of QR-sized Trojans by completely inelastic collisions of bodies
moving on co-planar orbits. An inelasticity parameter of
$f_{\rm KE} = 0.1$ is assumed for this figure; see equation (\ref{breakup}).
Curves are labelled by ($B$,$C$,$D$), where $B$ parameterizes
the semi-major axes of pre-collision bodies,
and $C$ and $D$ parameterize the pre-collision
eccentricities of masses $M$ and $m$, respectively.
Larger eccentricities and semi-major axes increasingly
different from $a_{\rm p}$
lead to reduced peak values of $f_{\rm col}$.
Curves for $(1,1,2)$, $(1,1.5,1.5)$, and $(1,2,1)$
are similar to but slightly below that for $(1,2,2)$ and not shown.
}
\label{colleps}
\end{figure}

\begin{figure}
\epsscale{0.8}
\plotone{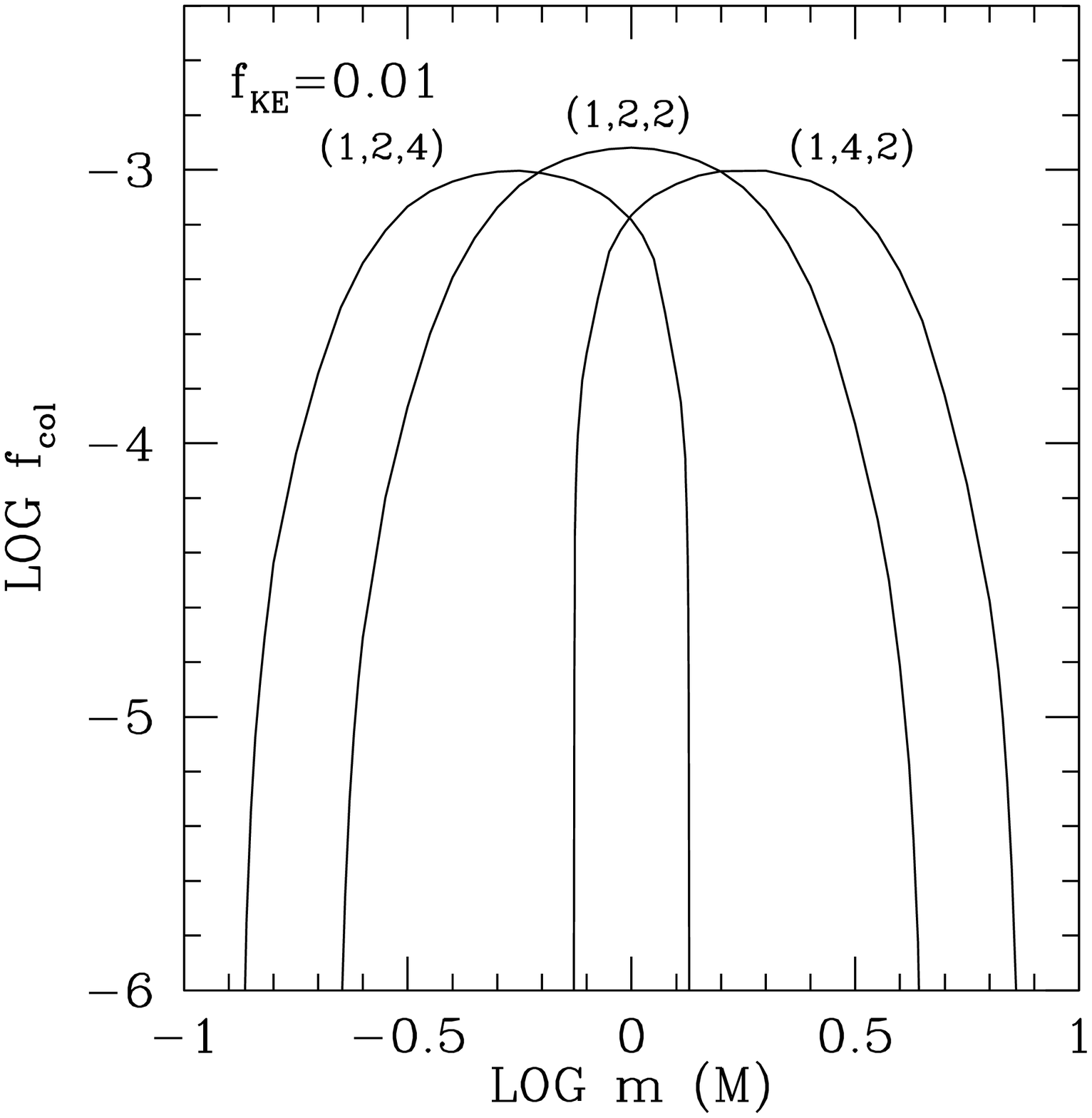}
\caption{Same as Figure \ref{colleps}, except
that $f_{\rm KE} = 0.01$, a value so low that
catastrophic dispersal is insignificant.
Efficiencies are higher than those for $f_{\rm KE} = 0.1$
and are symmetric about $m = M$.
}
\label{collepslowf}
\end{figure}

Families of curves for $f_{\rm col}$ as a function of $m$ 
are presented in Figures \ref{colleps} and \ref{collepslowf}.
The figures correspond to two different values of
$f_{\rm KE}$, 0.1 and 0.01, respectively.
Each curve is labelled by the parameter values ($B$,$C$,$D$).
The maximum efficiency attainable is $\max(f_{\rm col}) \sim 10^{-3}$,
appropriate for ($B$,$C$,$D$) = (1,2,2) and $m \approx M$.
Curves for $(1,1,2)$, $(1,1.5,1.5)$, and $(1,2,1)$
are similar to but still slightly below that for $(1,2,2)$ and not shown.

\placefigure{f5}

\placefigure{f6}

Successful collision geometries are those in which
the masses have orbital guiding centers (semi-major axes)
on opposite sides of L4. Typically one
mass is near the periapse of its orbit, while the other
is near apoapse. The maximum efficiency of $\sim$$10^{-3}$
can be rationalized as follows. From our computations,
the fraction of time
one mass spends near apoapse (at a potential Trojan-forming
position) is $\sim$$20^{\circ}/360^{\circ} \sim 0.055$.
The fraction of time the other spends near periapse
is similar, $\sim$0.055. Therefore given that a collision
has occurred, the probability that one mass was near periapse
and the other was at apoapse is $\sim$$2 \times (0.055)^2 \sim 0.006$.
The probability that the collision
occurred at the correct azimuth relative to Neptune,
in an arc of angular extent $\sim$1 rad centered on L4,
is $\sim$$1/2\pi$. Therefore the joint probability
is $0.006/2\pi \sim 10^{-3}$.

The reason why the efficiency curves for $D < C$ 
($e_m < e_M$) lie at $m > M$ (and vice versa)
can be understood by examining collisions that occur
exactly at periapse for one mass (say $m$) and exactly at apoapse
for the other ($M$): $a_m (1-e_m) = a_M (1+e_M) \approx a_{\rm p}$.
In a successful collision,
the post-collision velocity (now purely azimuthal) nearly equals
$v_{\rm p} = \Omega_{\rm p} a_{\rm p}$.
The pre-collision velocity of mass $m$ is
$v_m = (1+e_m)^{1/2} v_{\rm p}$, while
that of $M$ is $v_M = (1-e_M)^{1/2} v_{\rm p}$.
Success requires

\begin{equation}
M v_M + m v_m \approx (M+m) v_{\rm p} \,,
\end{equation}

\ni which we express as

\begin{equation}
\frac{m}{M+m} \approx \frac{ 1 - (1-e_M)^{1/2}}{ (1+e_m)^{1/2} - (1-e_M)^{1/2}
} \,.
\label{mmm}
\end{equation}

\ni Equation (\ref{mmm}) implies that $m \gtrsim M$ for
$e_m \lesssim e_M$. Similar conclusions obtain if $m$ and $M$
collide at their apoapse and periapse, respectively.

In Figure \ref{colleps}, for which $f_{\rm KE} = 0.1$,
efficiencies for $m > M$ are higher than those
for $m < M$ because for fixed $M$,
large projectile masses $m$ are more resistant
to catastrophic dispersal than small projectile masses: The left-hand-side
of equation (\ref{breakup2}) decreases as $(m/M)^{-5/3}$ for
$m > M$, but only as $m/M$ for $m < M$. This asymmetry is not
evident in Figure \ref{collepslowf}, for which
$f_{\rm KE} = 0.01$; catastrophic dispersal
is insignificant for such a high inelasticity.

Greater pre-collision eccentricities reduce peak values of $f_{\rm col}$
by producing greater relative velocities at impact;
these can either lead to catastrophic dispersal
or to Trojans having excessive eccentricities.

Removing the assumption of co-planarity
has the same effect as increasing pre-collision eccentricities. For example,
for ($B$,$C$,$D$) = (1,2,2), $m/M = 1$, and $f_{\rm KE} = 0.01$,
imposing a relative vertical velocity at the time of collision of
$0.10 \times \Omega_{m}a_{m}$, where $\Omega_{m}$ and $a_{m}$
are the mean motion and semi-major axis of mass $m$, respectively,
reduces $f_{\rm col}$ from the value shown in Figure \ref{collepslowf}
by a factor of 10. Imposing a relative
velocity of $0.15 \times \Omega_{m}a_{m}$ reduces
$f_{\rm col}$ to zero; the Trojans deposited
all have eccentricities $> 0.05$. Successful collisional emplacement
is rare for bodies having
pre-collision orbital planes that are misaligned by more
than $\sim$$6^{\circ}$.

While pre-collision orbital planes cannot have a mutual inclination
that is large, they still can be substantially
inclined with respect to the orbital plane of the planet.
Neptune Trojans enjoy dynamical stability at inclinations up to
$\sim$35$^{\circ}$ relative to Neptune's orbital plane
(Nesvorny \& Dones 2002).

\subsection{Final Requirements and Plausibility}
\label{upshot}

Armed with our appreciation for the underlying physics
of Neptune Trojan formation by direct collisional emplacement, we re-write
equation (\ref{coll2}) as

\begin{equation}
\Sigma_M \sim 0.2 \, \sigma_{\rm min} \left[ \frac{0.1t_{\rm sol}}{t_{\rm col}}
\, \frac{N_{\rm col}}{20} \, \frac{\max(f_{\rm col})}{f_{\rm col}}
\right]^{1/2} \left( \frac{R}{90\km} \right)^2\, ,
\label{finaleff}
\end{equation}

\ni where $\sigma_{\rm min} \sim 0.2 \gm/{\rm cm}^{2}$
is the surface density of solids in the minimum-mass nebula
at Neptune's heliocentric distance.
The maximum efficiency of $\max (f_{\rm col}) \sim 10^{-3}$
is attained for pre-collision bodies that orbit 
1--2 Neptune Hill radii from the planet and whose eccentricities
are 1--2 $\times$ $\mu^{1/3}$ $\sim 0.04$--$0.07$, i.e.,
marginally planet-crossing.

The requirements implied by equation (\ref{finaleff})---order unity
$\Sigma_{M}/\sigma_{\rm min}$
for maximum $f_{\rm col}$ and $t_{\rm col} \sim 4 \times 10^8\yr$---cannot
be met. 
The efficiency $f_{\rm col}$
cannot be maintained at its maximum value for such long $t_{\rm col}$.
Large objects within a few Neptune Hill radii of the planet
are on strongly chaotic orbits; their eccentricities random walk
to values of order unity on timescales much shorter than $\sim$$10^8$ yr.
Therefore $t_{\rm col} < 0.1 t_{\rm sol}$
and $f_{\rm col} < \max (f_{\rm col})$,
which imply that $\Sigma_{M}/\sigma_{\rm min} > 0.2$.
Such values of $\Sigma_{M}/\sigma_{\rm min}$
introduce a ``missing-mass'' problem (see, e.g.,
Morbidelli, Brown, \& Levison~2003).
Today, in the Kuiper belt at heliocentric distances
of $\sim$45 AU, $\Sigma_{M}/\sigma_{\rm min} \sim 10^{-3}$
(e.g., Bernstein et al.~2004), where $\Sigma_{M}$ is
interpreted as the surface density
of Kuiper belt objects (KBOs) having sizes $R \sim 90 \km$.
If $\Sigma_{M}/\sigma_{\rm min}$ were once greater than
order unity---as direct collisional
emplacement demands---how its value was thereafter reduced by
more than 3 orders of magnitude would need to be explained.
One resolution
to this problem posits that the number of bodies having
$R\sim 90$ km never exceeded its current low
value---that most of the condensates
in the local solar nebula instead accreted into
smaller, kilometer-sized objects (Kenyon \& Luu 1999).
Sub-kilometer-sized planetesimals are
favored by accretion models for Neptune for their high collision
rates and consequent low velocity dispersions (GLS04;
see also \S\ref{insitu}).

To summarize: To collisionally insert $N_{\rm col} \sim 20$ QR-sized objects
into libration about Neptune's L4 point requires a reservoir of
QR-sized objects having a surface density at least comparable
to and possibly greatly exceeding that of the minimum-mass disk of solids.
Neither observations of the Kuiper belt nor theoretical
models of planetary or KBO accretion support such a picture.
We therefore regard the formation of large Neptune Trojans
by direct collisional emplacement as implausible.

\section{FORMATION BY IN SITU ACCRETION}
\label{insitu}

While QR-sized objects are collisionally emplaced
with too low a probability,
much smaller objects---kilometer-sized planetesimals,
for example---may have deposited enough collisional debris into resonance
to build the current Trojan population (Shoemaker, Shoemaker,
\& Wolfe~1989; Ruskol 1990; Peale 1993).
Large Neptune Trojans present today
might have accreted {\it in situ} from such small-sized debris.

Modelling the collisional seeding process
would require that we understand the full spectrum of
sizes and orbital elements of pre-collision bodies, as well as the
size and velocity distributions of ejecta fragments.
Rather than embark on such a program,
we free ourselves from considerations
of the seeding mechanism and instead ask, given a seed population,
whether {\it in situ} accretion is viable at all.
We study the dynamics of growth inside the Trojan
resonance and quantify its efficiency
to constrain the surface density and radii of seed bodies,
the number of large Trojans that can form, and
the accretion timescale.
We will see that {\it in situ} accretion naturally reproduces
the observables with a minimal set of assumptions.

We adopt the two-groups approximation
[see, e.g., Goldreich et al.~2004 (GLS04)],
in which ``big'' bodies of radius $R$, mass $M$, and surface density $\Sigma$
accrete ``small'' bodies of radius $s$, mass $m$, and surface
density $\sigma \gtrsim \Sigma$. We define

\begin{equation}
g \equiv \sigma/\sigma_{\rm min}
\label{gmin1}
\end{equation}

\ni and

\begin{equation}
g_{\rm min} = \frac{2\pi N\rho R^3}{3\mu^{1/2}a_{\rm p}^2\sigma_{\rm min}}= 2
\times 10^{-4} \,.
\label{gmin2}
\end{equation}

\ni If $g = g_{\rm min}$, then
$\sigma$ is just sufficient to form $N = 20$
big bodies of radius $R=90\km$ out of the Trojan sub-disk of azimuthal
length $a_{\rm p}$ and radial width $2\mu^{1/2}a_{\rm p}$.
Note that $g = g_{\rm min}$ does not imply $\sigma = \sigma_{\rm min}$;
the surface density of the Trojan sub-disk may well have been
3 orders of magnitude lower than that of the local minimum-mass
disk of condensates.

Small bodies' epicyclic velocities, $u$,
are excited by viscous stirring from big bodies
and damped by inelastic collisions amongst small bodies.
Big bodies' epicyclic velocities, $v$,
are excited by viscous stirring from big bodies
and damped by dynamical friction with small bodies.
A characteristic velocity is

\begin{equation}
v_{\rm H} \equiv \Omega R_{\rm H} \,,
\end{equation}

\ni the Hill velocity from a big body,
where

\begin{equation}
R_{\rm H} = \left( \frac{M}{3M_{\odot}} \right)^{1/3} a = \frac{R}{\alpha}
\end{equation}

\ni is the Hill radius of a big body,

\begin{equation}
\alpha \equiv \left( \frac{3\rho_{\odot}}{\rho} \right)^{1/3}
\frac{R_{\odot}}{a} \approx 2 \times 10^{-4}
\end{equation}

\ni is a parameter
defined for convenience, and $\rho_{\odot}$ and $R_{\odot}$ are the average
density and radius of the Sun, respectively. Our analysis
draws heavily from the pedagogical review of planet
formation by GLS04, and the reader is referred there
regarding statements that we have not taken the space to prove here.

Any theory of {\it in situ} accretion must reproduce the observables,
$N \sim 10$--30 and $R \sim 90 \km$, within the age of
the Solar System. 
A promising guide is provided by the theory of dispersion-dominated,
oligarchic planet formation. Conventionally staged in a heliocentric
disk, the theory describes how each big body (``oligarch'')
gravitationally stirs and feeds from its own
annulus of radial half-width $\sim$$u/\Omega$, where
$v_{\rm esc} > u > v_{\rm H}$ and $v_{\rm esc} \sim v_{\rm H} / \alpha^{1/2}$
is the escape velocity from a big body.\footnote{We do
not develop shear-dominated ($u \lesssim v_{\rm H}$) Trojan oligarchy,
because we have discovered that it implicates, for $g$ not too far
above $g_{\rm min}$, seed particles so small that they
threaten to rapidly escape resonance by Poynting-Robertson drag.}
The dominance of each oligarch in its annulus
is maintained by runaway accretion.
We apply the theory of dispersion-dominated oligarchy to the Trojan sub-disk,
recognizing that annuli are now tadpole-shaped and centered on L4,
and that the
dynamics in the Trojan sub-disk are
more complicated than in an ordinary, heliocentric disk
because of the cycle of close-distant-close conjunctions (\S\ref{basic}).
We will need to juggle timescales such as $t_{\rm lib}$ and
$t_{\rm stir}$ that are absent in a non-Trojan environment.
Our purpose here is not to survey the entire range of
permitted accretion models but to explain how the simplest one works.
To this end, we will make assumptions
that simplify analysis and permit order-of-magnitude estimation.
Many of these assumptions we will justify in \S\S\ref{check1}--\ref{check3}.
Those that we do not are listed in \S\ref{uine}.


We derive $u$ by balancing viscous stirring by big bodies
against inelastic collisions with small bodies.
We work in the regime where the time between
collisions of small bodies,

\begin{equation}
t_{\rm col,u} \equiv - u \left. \frac{du}{dt}\right|_{\rm col}^{-1} \sim
\frac{\rho s}{\sigma \Omega} \,,
\end{equation}

\ni exceeds the grand synodic time, $t'_{\rm syn,sl}$, between
a typical small body and its
nearest neighboring big body (see \S\ref{anharm}).
This choice, which essentially places a lower bound
on the small-body sizes $s$ that we consider,
is made so that we may employ simple time-averaged
expressions for various rates.
We evaluate $t'_{\rm syn,sl}$
at $\delta a = 0.5\mu^{1/2}a$ and $x = u/\Omega$ [see
equation (\ref{gst})].

The timescale, $t_{\rm vs,u}$, for viscous stirring to double $u$ is the
mean time between close conjunctions of a small body with its nearest
neighboring big body, multiplied by the number of conjunctions
required for $u$ to random walk to twice its value.
Each conjunction changes $u$ by
$\Delta u \sim [GM/(u/\Omega)^2] \Omega^{-1} \sim v_{\rm H}^3/u^2$.
The number of conjunctions required for $u$ to double
is $(u/\Delta u)^2 \sim (u/v_{\rm H})^6$.
Since close conjunctions occur at the time-averaged rate of
$(t_{\rm stir}/t_{\rm syn})/t'_{\rm syn,sl} \sim u/a$,
the timescale for viscous stirring to double $u$ is

\begin{equation}
t_{\rm vs,u} \equiv u \left. \frac{du}{dt} \right|_{\rm vs}^{-1} \sim \left(
\frac{u}{v_{\rm H}} \right)^6 \frac{a}{u} \,.
\label{superhill}
\end{equation}

\ni Equating $t_{\rm vs,u}$ with $t_{\rm col,u}$ gives the equilibrium
velocity

\begin{equation}
\frac{u}{v_{\rm H}} \sim \left( \frac{v_{\rm H} t_{\rm col,u}}{a}
\right)^{1/5}\,,
\label{uvh}
\end{equation}

\ni valid for $t_{\rm col,u} \gtrsim t'_{\rm syn,sl}$.
The condition $t_{\rm col,u} = t'_{\rm syn,sl}$
implies that 

\begin{equation}
\frac{u}{v_{\rm H}} \approx 43^{1/6} \approx 1.9\,,
\label{minu}
\end{equation}

\ni independent of $R$ and $g$. We adopt
this value for $u/v_{\rm H}$ in the remaining discussion.
We have assumed that $u$ approximates
the relative velocity between small and big bodies during a conjunction;
dynamical friction cooling of big bodies by small bodies ensures that
this is the case, as shown
in \S\ref{check1}.

How oligarchic accretion ends determines the final radius, $R_{\rm final}$,
of a big body. Oligarchy might end when $\Sigma \sim \sigma$.
At this stage, big bodies undergo a velocity
instability in which viscous stirring overwhelms
cooling by dynamical friction
and $v$ runs away (GLS04; \S\ref{postdate}; see also \S\ref{check1}).
Larger relative velocities weaken gravitational focussing
and slow further growth of big bodies.

There is, however, another way in which 
oligarchic accretion can end in the Trojan sub-disk: collisional
diffusion of small bodies out of resonance. 
Small bodies can random walk out of the sub-disk before
big bodies can accrete them to the point when $\Sigma \sim \sigma$
(where $\sigma$ is understood as the original surface
density of small bodies, evaluated before loss by diffusion is appreciable).
We assume that loss by diffusion halts accretion and check our assumption
in \S\ref{check2}. Changes in the libration amplitudes of big bodies are
ignorable, as shown in \S\ref{check3}.
The diffusion time of small bodies is estimated as follows.
The orbital guiding centers of small bodies shift
radially by about $\pm u/\Omega$ every $t_{\rm col, u}$.
Small bodies random walk out of the resonance over a timescale

\begin{equation}
t_{\rm esc,s} \sim t_{\rm col,u} \left( \frac{\mu^{1/2}a}{u/\Omega} \right)^2
\,,
\end{equation}

\ni which for our assumption that $t_{\rm col,u} = t'_{\rm syn,sl}$ equals

\begin{equation}
t_{\rm esc,s} \sim 43 \left( \frac{\alpha a}{R} \right)^3 \left( \frac{v_{\rm
H}}{u} \right)^3 \frac{\mu}{\Omega}\,.
\end{equation}

\ni 
We equate $t_{\rm esc,s}$ to the assembly time of a big body, $t_{\rm acc}$,
to solve for $R_{\rm final}$.
Accretion of small bodies to form a big body is accelerated
by gravitational focussing, whence\footnote{Runaway accretion is embodied in
equation (\ref{tacc}).
Consider two oligarchs having radii $R$ and $\tilde{R}$.
If the excitation/feeding annuli of the two oligarchs
overlap, the bigger oligarch out-accretes
the smaller, since $t_{\rm acc}/\tilde{t}_{\rm acc} \sim (R/v_{\rm
H}^2)/(\tilde{R}/\tilde{v}_{\rm H}^2) \sim \tilde{R}/R$.
This scaling is independent of $u$ and $\sigma$ because those
variables are common to both competing oligarchs.}

\begin{equation}
t_{\rm acc} \sim \frac{\rho R}{\sigma \Omega} \left( \frac{u}{v_{\rm esc}}
\right)^2 \sim \frac{\rho R \alpha}{\sigma \Omega} \left( \frac{u}{v_{\rm H}}
\right)^2 \,.
\label{tacc}
\end{equation}

\ni Equating $t_{\rm esc,s}$ to $t_{\rm acc}$ yields

\begin{equation}
R = R_{\rm final} \sim \left( \frac{43 \mu \alpha^2 a^3 \sigma}{\rho}
\right)^{1/4} \left( \frac{v_{\rm H}}{u} \right)^{5/4} \sim 150 \km \left(
\frac{g}{10g_{\rm min}} \right)^{1/4} \,.
\label{racc}
\end{equation}

\ni For $R \sim R_{\rm final}$, it follows that

\begin{equation}
t_{\rm acc} \sim 1 \times 10^9 \left( \frac{10g_{\rm min}}{g} \right)^{3/4} \yr
\,,
\label{finalt}
\end{equation}
\begin{equation}
u  \sim  1.9 v_{\rm H}  \sim  2 \left( \frac{g}{10g_{\rm min}} \right)^{1/4} \m
\, \s^{-1} \,,
\end{equation}
\begin{equation} \label{smallbody}
s  \sim  20 \left( \frac{g}{10g_{\rm min}} \right)^{3/4} \cm \,,
\end{equation}

\ni and that the number of oligarchs accreted equals

\begin{equation}
N_{\rm acc} \sim \frac{\mu^{1/2}a}{2u/\Omega} \sim 10 \left( \frac{10g_{\rm
min}}{g} \right)^{1/4} \,.
\label{naccfinal}
\end{equation}

\ni We regard (\ref{racc}), (\ref{finalt}), and (\ref{naccfinal})
as plausibly explaining the observations: Accretion of deci-meter-sized
particles having a surface density $\sim$10 times that present
in QR-sized bodies today generates $\sim$10 large Neptune
Trojans in $\sim$1 Gyr. Their growth naturally halts
from the diffusive escape of small bodies out of resonance.
Large Neptune Trojans may well be among
the most recently assembled bodies in the Solar System.

\subsection{Checking $v < u$}
\label{check1}

We have assumed and now check that $v < u$ so that
$u$ approximates the relative velocity between small and big
bodies during a conjunction.
Big bodies cool by dynamical friction with small bodies.
Since a big body continuously
undergoes conjunctions with small bodies in the same manner
that it would outside resonance (see \S\ref{trojtroj}), the standard formula
(GLS04) for dynamical friction cooling of big bodies applies:

\begin{equation}
t_{\rm df,v} \equiv -v \left. \frac{dv}{dt} \right|_{\rm df}^{-1} \sim
\frac{\rho R \alpha^2}{\sigma \Omega} \left( \frac{u}{v_{\rm H}} \right)^4 \sim
7 \times 10^5 \left( \frac{R}{R_{\rm final}} \right) \left( \frac{10g_{\rm
min}}{g} \right) \left( \frac{u}{1.9v_{\rm H}} \right)^4 \yr\,.
\label{tdf}
\end{equation}
 
\ni To solve for $v$, we set this cooling timescale equal to
$t_{\rm vs,v}$, the timescale
for viscous stirring by big bodies to double $v$.
We exploit the fact that for $R \sim R_{\rm final}$,
$t_{\rm df,v}$ is of order
$t'_{\rm syn,ll} \sim 1 \times 10^6\,(R_{\rm final} / R) \yr$
(the grand synodic time between neighboring big bodies
separated by $x = 2 u/\Omega$) to
derive $t_{\rm vs,v}$ in the same way that we derived
$t_{\rm vs,u}$.\footnote{When $R \ll R_{\rm final}$, 
$t_{\rm df,v} \ll t'_{\rm syn,ll}$ and
$v < v_{\rm H} < u$ whilst the big body is unstirred by its
nearest big neighbor, which is the majority of the time.}
The key simplification in that derivation was our
ability to time-average over the cycle of close-distant-close conjunctions.
Each close conjunction between neighboring oligarchs
imparts $\Delta v \sim \Delta u / 4\sim v_{\rm H}^3 / 4 u^2$.
The number of conjunctions required to double $v$
is $(v/\Delta v)^2 \sim 16 v^2 u^4 / v_{\rm H}^6$.\footnote{Our estimate
is valid if
$\Delta v < v$, which in turn demands that $u/v_{\rm H} \gtrsim (\pi/6)^{1/6}
(\sigma/\Sigma)^{1/6}$. Physically this means that a given big body is stirred
primarily by its nearest big neighbors. This inequality is satisfied for times
near $t_{\rm acc} \sim t_{\rm esc,s}$.\label{rsari}}
Since the time-averaged rate of close conjunctions between neighboring
big bodies is $(t_{\rm stir}/t_{\rm syn})/t'_{\rm syn,ll} \sim 2u/a$,

\begin{equation}
t_{\rm vs,v} \equiv v \left. \frac{dv}{dt} \right|_{\rm vs}^{-1} \sim 8
\frac{v^2 u^4}{v_{\rm H}^6} \frac{a}{u} \,.
\end{equation}

\ni We cast this expression into a form more closely resembling
equation (\ref{tdf}) by noting that the surface density
of big bodies in dispersion-dominated oligarchy is

\begin{equation}
\Sigma \sim \frac{M}{4 a (u/\Omega)} \,,
\end{equation}

\ni since each big body occupies a feeding annulus of width
$\sim$$2u/\Omega$ and perimeter $\sim$$2a$. Then

\begin{equation}
t_{\rm vs,v} \sim \frac{8\pi}{3} \frac{\rho R \alpha^2}{\Sigma \Omega}
\frac{v^2 u^2}{v_{\rm H}^4} \,.
\label{myformula}
\end{equation}

\ni Equating $t_{\rm vs,v}$ with $t_{\rm df,v}$ yields

\begin{equation}
\frac{v}{u} \sim \left( \frac{3}{8\pi} \frac{\Sigma}{\sigma} \right)^{1/2} \,.
\label{vu}
\end{equation}

\ni Therefore $v < u$ provided $\Sigma \lesssim (8\pi/3)\, \sigma$;
this inequality is well satisfied while big Trojans grow.
Goldreich et al.~(2004, GLS04) point out that if $\Sigma \gtrsim \sigma$,
dynamical friction cooling fails to balance viscous stirring
and $v$ de-stabilizes. Though we agree (to order-of-magnitude) with
this conclusion, the power-law index in 
equation (\ref{vu}) should be 1/2, not 1/4 as
stated in equation (109) of GLS04.
The error arises in GLS04 because these authors assume
that epicycles of big bodies overlap;
in dispersion-dominated Trojan oligarchy for $v < u$,
they do not.

\subsection{Checking that Diffusive Loss Limits Accretion}
\label{check2}

Our assumption that loss by diffusion of small bodies limits accretion
is valid if $R_{\rm final} < R_{\rm iso,Trojan}$,
where the latter radius is that of the isolation mass.
In dispersion-dominated oligarchy, the isolation mass is that contained in a
tadpole-shaped annulus
of perimeter $\sim$$2a_{\rm p}$ and radial width $\sim$$2u/\Omega$:

\begin{equation}
\frac{4\pi \rho R_{\rm iso,Trojan}^3}{3} \sim \frac{4 a_{\rm p} u
\sigma}{\Omega} \,,
\end{equation}

\ni from which we derive

\begin{equation}
R_{\rm iso,Trojan} \sim \left( \frac{3 a_{\rm p} \sigma u}{\pi \rho \alpha
v_{\rm H}} \right)^{1/2} \sim 300 \left( \frac{g}{10g_{\rm min}} \right)^{1/2}
\km > R_{\rm final}\,.
\label{risotrojan2}
\end{equation}

\ni Thus escape of small bodies by diffusion
prevents big bodies from undergoing their last potential radius-doubling.
Velocity instability still eventually occurs; but it is triggered
by decay of $\sigma$ by escape of small bodies, not
by accretion of big bodies.

\subsection{Checking that Big Bodies Do Not Migrate}
\label{check3}

We have assumed that the libration amplitudes
of big bodies do not change significantly over the
accretion epoch. First we consider how big bodies
change the libration amplitude of a big body; the
effects of small bodies are considered separately
in the latter half of this sub-section.
A lower bound on the time it takes
a big body to escape the resonance can be
established by considering the big bodies to be spaced by $x = R_{\rm H}$.
In this case, the semi-major axis of a big body
changes by $\pm R_{\rm H}$ every time the body
undergoes a close conjunction with its nearest neighbor (in other words,
the big bodies swap places every close conjunction).
Then a big body random walks out of the resonance
over a timescale of $t_{\rm diff,b} \sim (\mu^{1/2} a / R_{\rm H} )^2
a / v_{\rm H} \sim 2 \times 10^8 \, (150 \km / R)^3 \yr$, where
$\sim$$a/v_{\rm H}$ is the time between close conjunctions,
time-averaged over the grand synodic period. Though this diffusion
time is a factor of 5 shorter than
$t_{\rm acc} \sim t_{\rm esc,s} \sim 10^9 \yr$,
the true escape time of big bodies will be larger than
$t_{\rm diff,b}$ because the actual epicycles traced by big
bodies---and therefore the characteristic stepsizes in any
random walk---only approach $\sim$$R_{\rm H}/20$ in size.
Crude estimates suggest that the true escape time of big bodies
due to interactions with other big bodies
is orders of magnitude longer than the above estimate
of $t_{\rm diff,b}$ for our model parameters.

Viscous stirring by small bodies is much less effective than
viscous stirring by big bodies. Since the epicycles of small bodies cross
the orbits of big bodies, the timescale for viscous stirring
by small bodies to double $v$ is given by the usual formula

\begin{equation}
\left. t \right|_{\rm vs,v,s} \sim \frac{\rho R \alpha^2}{\sigma \Omega}
\frac{Mv^2}{mu^2} \frac{u^4}{v_{\rm H}^4}
\end{equation}

\ni (GLS04). This timescale is safely
longer than that for viscous stirring by
big bodies to double $v$ [equation (\ref{myformula})] by

\begin{equation}
\frac{ \left. t \right|_{\rm vs,v,s}}{ \left. t \right|_{\rm vs,v}} \sim
\frac{3}{8\pi} \frac{\Sigma}{\sigma} \frac{M}{m} \gg 1\,.
\end{equation}

Changes in libration amplitudes of big bodies
by dynamical friction cooling by small bodies are also
ignorable.
A big body experiences friction with small bodies inside and outside
its tadpole-shaped, guiding center orbit. Interactions with
the interior disk of small bodies will be of the
same magnitude as interactions with the exterior disk, differing
only by of order $\epsilon \ll 1$. 
Simple estimates suggest that $\epsilon < 10^{-2}$.
If the surface density of small bodies changes fractionally
by order unity from the Lagrange point to the outer edge of the resonance,
then $\epsilon \sim (v/\Omega) / (\mu^{1/2} a) \sim 5 \times 10^{-3}$.
If no such large scale gradient exists,
$\epsilon \sim v/(\Omega a) \sim 4 \times 10^{-5}$.
During each period of distant
conjunctions, when small bodies reduce the size of the big
body's epicycle by of order $v/\Omega$
over the cooling timescale $t_{\rm df,v}$,
the change to the size of the tadpole orbit will be at most of order
$|\Delta (\delta a)| \sim \epsilon (v/\Omega)$. Conservatively we assume
that the change always has the same sign; that is, the big body
suffers a steady drift either towards or away from the Lagrange point
at a velocity

\begin{equation}
w_{\rm drift,b} \sim \frac{ \epsilon v/\Omega }{t_{\rm df,v}} \,.
\end{equation}

\ni The big body migrates across the width of the resonance over a time

\begin{equation}
t_{\rm drift,b} \sim \frac{\mu^{1/2}a}{w_{\rm drift,b}} \sim 1 \times 10^{10}
\left( \frac{0.01}{\epsilon} \right) \yr
\end{equation}

\ni where we have used $v \sim u/10$, $u \sim 2 \m/\s$, and
$t_{\rm df,v} \sim 7 \times 10^5 \yr$. Therefore unless
$\epsilon \gg 10^{-2}$, systematic drifts
of a big body can be ignored over
$t_{\rm acc} \sim t_{\rm esc,s} \sim 10^9 \yr$.

\subsection{Neglected Effects and Unresolved Issues}
\label{uine}

In estimating $u$ and $v$, we assumed that
the dominant source of velocity excitation during
the era of accretion was viscous
stirring by large Trojans.
Our neglect of velocity excitation by the other giant planets
during this early epoch remains to be justified.

The epicyclic velocities, $v$, of big bodies
considered above
do not exceed $u \sim 2$ m/s, yet today
QR's epicyclic velocity is $\sim$100 m/s.
Can viscous stirring, unchecked by dynamical friction after
the onset of the velocity instability, produce such a large
$v$? The answer is no; $v \sim 100$ m/s is of order
$v_{\rm esc}$ and producing it would require near-grazing collisions between
QR-sized bodies for which the timescale
is $\sim$$t_{\rm life} \sim 7 \times 10^{13}\yr$
(see \S\ref{pastpop}). Therefore we appeal instead to
velocity excitation by the other giant planets, occurring after
the velocity instability, to generate the epicyclic amplitudes
observed today. Perturbations due to the $\nu_{18}$ inclination resonance
and other secular resonances (Marzari et al.~2003;
Brasser et al.~2004) seem adequate to the task.
Our speculation should be straightforward
to verify with numerical orbit integrations.

Finally, the two-groups analysis assumes
that the sizes of small bodies do not change while
big bodies accrete.
Future accretion models should incorporate
a spectrum of planetesimal sizes to test this assumption.

\section{SUMMARY AND DIRECTIONS FOR FUTURE WORK}
\label{conclude}

Neptune Trojans represent one of the most recent
additions to the Solar System's inventory of dynamical
species [Chiang et al.~2003 (C03)]. In this work, we have systematically
reviewed their properties by combining observation with theory.
We have further assessed how three theories---pull-down capture,
direct collisional emplacement, and {\it in situ} accretion---fare
in explaining these properties. 
We wish to elucidate the genesis
of Neptune Trojans not only to understand this new
class of object in and of itself, but also to shed light
on the still debated circumstances of formation of the host ice giant
[Thommes et al.~1999; C03; Goldreich et al.~2004 (GLS04)].
Insights into the accretion of Trojans in disks surrounding
Lagrange points can carry over
into the accretion of planets in disks surrounding stars.

We summarize our main findings as follows.

\begin{enumerate}
\item{Between $\sim$10 and $\sim$30
objects comparable in size to the Neptune Trojan 2001 QR$_{322}$ (QR)
are trapped near Neptune's forward Lagrange (L4) point.
Presumably a similar population exists at L5.
Numerical orbit integrations suggest these objects
have occupied the 1:1 resonance
for the age of the Solar System, $t_{\rm sol}$.
The characteristic radius of QR ($R \sim 90\km$) is comparable to that
of the largest known Jupiter Trojan, (624) Hektor. As a function
of size for $R \gtrsim 65 \km$, Neptune Trojans are
at least as numerous as their Jovian
counterparts and may outnumber them by a factor of $\sim$10--30.
Their surface mass density approaches that of the main Kuiper belt
today to within a factor of a few, and is 3--4 orders of magnitude
lower than that of the local minimum-mass disk of condensates.}

\item{The number of large (QR-sized) Neptune Trojans has not been altered
by destructive, intra-cloud collisions. Lifetimes
against catastrophic dispersal exceed $t_{\rm sol}$ by more
than 2 orders of magnitude. That large Neptune
Trojans are today essentially collisionless suggests that
their number directly reflects the efficiency of a primordial
formation mechanism.}

\item{The existence of Neptune Trojans can be reconciled with
violent orbital histories of Neptune (Thommes et al.~1999; GLS04),
provided that Trojans form after
the planet has its orbit circularized. In histories where
Neptune's eccentricity was once of order unity,
final circularization is achieved by
dynamical friction with ambient planetesimals.
Long-term occupancy of the Trojan resonance is
possible after circularization. This point was not recognized
by C03. We regard their argument that present-day Neptune
Trojans rule out dramatic scattering events for Neptune to
be overstated.}

\item{Pull-down capture, whereby objects are trapped into
Trojan resonance by mass growth of the host planet,
is unlikely to have been solely responsible for the origin
of large Neptune Trojans. Libration amplitudes
(i.e., tadpole orbit sizes) are damped by pull-down capture too inefficiently
to explain the libration amplitude
of QR. Moreover, the present theory is unsatisfactory because
it assumes, without explanation, the prior existence
of objects on 1:1 horseshoe orbits. The theory
fails to specify what determines
the number of weakly bound librators
at the onset of pull-down capture.
}

\item{A pair of initially non-resonant, QR-sized objects can be diverted onto
low-eccentricity Trojan orbits
(like that occupied by QR) when they collide near the periapse
and apoapse of their respective orbits.
The greatest efficiencies of direct collisional emplacement
are attained for planetesimals that orbit 1--2 Hill
radii away from the planet and whose eccentricities are
1--2 $\times$ $\mu^{1/3}$ $\sim 0.04$--$0.07$,
where $\mu = 5 \times 10^{-5}$ is the ratio of Neptune's
mass to the Sun's. Such orbits are strongly chaotic
and cannot be maintained over collision timescales.
We estimate that actual efficiencies of direct
emplacement are so low that
a heliocentric disk of QR-like planetesimals can divert
$N_{\rm col} \sim 20$ of its members into
resonance only if the disk's surface density
exceeds $\sim$20\% that of the local minimum-mass disk of solids.
Unfortunately, such a large surface density in QR-sized objects
is not supported by observations of the Kuiper belt or
by theoretical models of how Neptune and Kuiper belt
objects accreted. We therefore discount the formation of large
Neptune Trojans by direct collisional emplacement.}

\item{{\it In situ} accretion of large Neptune Trojans
is viable and attractive. We exercised the two-groups approximation
to study accretion dynamics within a primordial
Trojan sub-disk composed of small
seed bodies having sizes $s \sim 20 \cm$ and a surface
density lower than that of the local
minimum-mass disk of condensates by $g \sim 2 \times 10^{-3}$.
This surface density is 10 times
that of the Neptune Trojan sub-disk today (in QR-sized objects).
A plausible way to seed the resonance is by
planetesimal collisions that insert ejecta fragments
into the Trojan libration region. Big bodies accrete small
bodies in our model sub-disk to grow to a radius of
$R_{\rm final} \sim 150 \km$ over a period
of $t_{\rm acc} \sim 1 \times 10^9 \yr$.
Their number at this time is $N_{\rm acc} \sim 10$, as mandated
by the rules of dispersion-dominated oligarchy.
Collisional diffusion of small bodies out of the resonance
naturally halts further growth. Large Neptune Trojans may be
the most recently assembled objects in our planetary system.}

\end{enumerate}

We have developed the case that the number and sizes of
large Neptune Trojans represent an unadulterated
imprint of oligarchic accretion inside resonance.
Confirmation of this result would support theories
of planet formation by accretion of very small particles (GLS04).
What future observations or theoretical work
might help to develop this picture?
Measurement and calculation of the size distribution
of Trojans are natural next steps. Perhaps more intriguing
is the question of velocity dispersions inside
the Neptune Trojan cloud.

Today, the eccentricity and inclination (with respect
to the invariable plane) of QR are both about 0.03.
The corresponding epicyclic velocities are of order 200 m/s.
In contrast, our accretion model requires velocity dispersions $\lesssim$ 2 m/s
while big bodies grow.
We suspect, but have yet to check, that today's velocity dispersions
are the result of gravitational perturbations
exerted by the other giant planets---perturbations
unchecked by dynamical friction during the
present, non-accretionary epoch.
We might expect planetary perturbations to amplify
eccentricities and inclinations of Neptune Trojans to values
not exceeding $\sim$0.1. This expectation stems from billion-year-long
integrations of possible trajectories of QR (Brasser et al.~2004),
which reveal that its eccentricity and inclination (with respect
to the invariable plane) stay below $\sim$0.1.
For some trajectories, the $\nu_{18}$
resonance is found to raise the inclination
of QR to at most $7^{\circ}$ (0.12 rad).
More typically, the inclination
remains below $\sim$$1.6^{\circ}$ (Brasser et al.~2004).
Thus, even though Neptune
Trojans can exist at inclinations as high as $\sim$$35^{\circ}$
(Nesvorny \& Dones 2002; Marzari et al.~2003),
we see no reason why these niches at high inclination
should be occupied. We might expect mutual inclinations between
large Neptune Trojans to be less than $\sim$$10^{\circ}$.
This picture of a ``thin disk'' contrasts
with the ``thick disks'' of other
minor body belts---Jupiter Trojans, Main Belt asteroids,
and the Kuiper belt\footnote{With the possible exception
of the ``core'' Classical Kuiper belt having inclinations
less than $\sim$4.6 degrees with respect to their mean plane
(Elliot et al.~2005). Like Neptune Trojans, this cold population
may also be relatively dynamically pristine.}---and
reflects the dynamically
cold accretionary environment that we have championed.
It is subject to observational test.


\acknowledgements
We thank D.~Jewitt, S.~Kortenkamp,
H.~Levison, J.~Lovering, R.~Malhotra, R.~Murray-Clay, R.~Sari,
and the DES collaboration for helpful discussions. Exchanges with D.J.
and H.L.~regarding difficulties with pull-down
capture were especially motivating.
Footnote \ref{rsari} is due to R.~Sari. Simulations of divergent resonance
crossing by R.~Murray-Clay were helpful in writing the introduction.
Our use of the verb ``to anneal'' in the abstract is inspired
by R.~Malhotra. We thank the referee, P.~Goldreich, for a
prompt and thought-provoking report that helped to improve
the presentation of this paper.
This work was supported by the National Science
Foundation and the Alfred P.~Sloan Foundation.



\begin{references}
Barucci, M.A., Cruikshank, D.P., Mottola, S., \& Lazzarin, M. 2002, in
Asteroids III, eds. W.F. Bottke, Jr., A. Cellino, P. Paolicchi, \& R.P. Binzel
(Tucson: Univ. Arizona Press), 273

Bernstein, G., \& Khushalani, B.~2000, \aj, 120, 3323

Bernstein, G., et al.~2004, \aj, 128, 1364

Brasser, R., et al.~2004, \mnras, 347, 833

Chiang, E.I., \& Jordan, A.B.~2002, \aj, 124, 3430

Chiang, E.I.~2003, \apj, 584, 465

Chiang, E.I., et al.~2003, \aj, 126, 430 (C03)

Davis, D.R., et al.~2002, in Asteroids III, eds. W.F. Bottke, Jr., A. Cellino,
P. Paolicchi, \& R.P. Binzel (Tucson: Univ. Arizona Press), 545


Elliot, J.L., et al.~2005, \aj, 129, 1117

Fernandez, J.A., \& Ip, W.H.~1984, Icarus, 58, 109

Fleming, H.J., \& Hamilton, D.P.~2000, Icarus, 148, 479

Goldreich, P., Lithwick, Y., \& Sari, R.~2004, \araa, 614, 497 (GLS04)

Gomes, R.~1998, \aj, 116, 2590

Gomes, R.~2001, \aj, 122, 3485

Gomes, R.~2003, Icarus, 161, 404

Greenberg, R., Bottke, W.F., Carusi, A., \& Valsecchi, G.B.~1991, Icarus, 94,
98

Hahn, J.M., \& Malhotra, R.~1999, \aj, 117, 304

Hartmann, W.K., \& Cruikshank, D.P.~1978, Icarus, 36, 353

Heppenheimer, T.A., \& Porco, C.~1977, Icarus, 30, 385

Holman, M., \& Wisdom, J.~1993, \aj, 105, 1987

Holsapple, K., et al.~2002, in Asteroids III, eds. W.F. Bottke, Jr., A.
Cellino, P. Paolicchi, \& R.P. Binzel (Tucson: Univ. Arizona Press), 443

Jewitt, D.C., Sheppard, S., \& Porco, C.~2004, in Jupiter. The planet,
satellites and magnetosphere, eds. F.~Bagenal, T.E.~Dowling, \& W.B.~McKinnon
(Cambridge: University Press), 263

Jewitt, D.C., Trujillo, C.A., \& Luu, J.X.~2000, \aj, 120, 1140

Karlsson, O.~2004, Astron. \& Astrophys., 413, 1153

Kenyon, S.J., \& Luu, J.X.~1999, \aj, 118, 1101

Kortenkamp, S.J., Malhotra, R., \& Michtchenko, T.~2004, Icarus, 167, 347

Lagrange, J.-L., 1873, in Oeuvres de Lagrange, Tome Sixi\`{e}me, ed. J.-A.
Serret (Paris: Gauthiers-Villars), 229

Lodders, K., \& Fegler, B.~1998, The Planetary Scientist's Companion (New York:
Oxford University Press)

Malhotra, R.~1995, \aj, 110, 420

Marzari, F., \& Scholl, H.~1998a, Icarus, 131, 41

Marzari, F., \& Scholl, H.~1998b, Astron.~\& Astrophys., 339, 278

Marzari, F., Scholl, H., Murray, C., \& Lagerkvist, C. 2002, in Asteroids III,
eds. W.F. Bottke, Jr., A. Cellino, P. Paolicchi, \& R.P. Binzel (Tucson: Univ.
Arizona Press), 725

Marzari, F., Tricarico, P., \& Scholl, H.~2003, Astron.~\& Astrophys., 410,
725

Matsuyama, I., Johnstone, D., \& Hartmann, L.~2003, \apj, 582, 893

Morbidelli, A., Brown, M.E., \& Levison, H.F.~2003, Earth, Moon, \& Planets 92,
1

Morbidelli, A., Levison, H.F., Tsiganis, K., \& Gomes, R., AAS DPS meeting
\#36, abstract \#40.03

Murray, C.D., \& Dermott, S.F.~1999, Solar System Dynamics (New York: Cambridge
University Press)

Murray-Clay, R.A., \& Chiang, E.I.~2005, \apj, 619, 623

Nesvorny, D., \& Dones, L.~2002, Icarus, 160, 271

Peale, S.J.~1993, Icarus, 106, 308

Press, W.H., Teukolsky, S.A., Vetterling, W.T., \& Flannery, B.P.~1992,
Numerical Recipes in Fortran.~The Art of Scientific Computing (Cambridge:
University Press)

Ruskol, E.L.~1990, Astron.~Vestnik, 24, 244 [in Russian]

Shoemaker, E.M., Shoemaker, C.S., \& Wolfe, R.F.~1989, in Asteroids II, eds.
R.P. Binzel, T. Gehrels, \& M.S. Matthews (Tucson: Univ. Arizona Press), 487

Tanga, P., et al.~2003, A\&A, 401, 733

Thommes, E.W., Duncan, M.J., \& Levison, H.F.~1999, Nature, 402, 635
\end{references}
\end{document}